\documentclass[aps,twocolumn,superscriptaddress,showpacs]{revtex4}
\usepackage{amsmath}
\usepackage{amssymb}
\usepackage{amsfonts}
\usepackage{graphicx}
\usepackage{bm}

\usepackage{dcolumn}
\usepackage{bm}
\usepackage{graphicx}
\usepackage{color}
\DeclareGraphicsExtensions{. pdf, . mps, . png, . eps, . ps, . EPS}

\begin{document}
\def\be{\begin{equation}}
\def\ee{\end{equation}}
\def\bc{\begin{center}} 
\def\ec{\end{center}}
\def\bea{\begin{eqnarray}}
\def\eea{\end{eqnarray}}
\newcommand{\avg}[1]{\langle{#1}\rangle}
\newcommand{\Avg}[1]{\left\langle{#1}\right\rangle}

\title{Social network dynamics of face-to-face interactions}

\author{Kun Zhao}
\affiliation{Department of Physics, Northeastern University, Boston 02115 MA, USA}
\author{Juliette Stehl\'e}
\affiliation{Centre de Physique Th\'eorique, CNRS (UMR 6207) et Universit\'e d'Aix-Marseille, 
Campus de Luminy, Case 907, F-13288 Marseille cedex 9,
France}
\author{ Ginestra Bianconi}  
\affiliation{Department of Physics, Northeastern University, Boston 02115 MA, USA}
\author{Alain Barrat}
\affiliation{Centre de Physique Th\'eorique, CNRS (UMR 6207) et Universit\'e d'Aix-Marseille,
Campus de Luminy, Case 907, F-13288 Marseille cedex 9,
France}
\affiliation{Complex Networks and Systems Group,
Institute for Scientific Interchange (ISI), Torino 10133, Italy}

\begin{abstract}
The recent availability of data describing social networks is changing
our understanding of the ''microscopic structure" of a social tie.  A
social tie indeed is an aggregated outcome of many social interactions
such as face-to-face conversations or phone calls. Analysis of data on
face-to-face interactions shows that such events, as many other human
activities, are bursty, with very heterogeneous durations.  In this
paper we present a model for social interactions at short time scales, 
aimed at describing contexts such as conference venues in which individuals
interact in small groups. We present a detailed analytical and numerical
study of the model's dynamical properties, and show that it reproduces
important features of empirical data. The model
allows for many generalizations toward an increasingly realistic
description of social interactions. In particular in
this paper we investigate the case where the agents have intrinsic
heterogeneities in their social behavior, or where dynamic variations
of the local number of individuals are included. Finally we propose
this model as a very flexible framework to investigate how dynamical
processes unfold in social networks.
\end{abstract}
\maketitle

\section{Introduction}

In the last decade complexity theory has  greatly advanced thanks 
to the availability of extensive data on a wide variety of networked
systems.
Empirical studies have uncovered the presence of ubiquitous features
in complex networks, such as the small-world property, or strong
heterogeneities in the topological structure, revealed for instance by
broad degree distributions 
\cite{Dorogovtsev:2003,Newman:2003,Pastor:2004,Boccaletti:2006,Caldarelli:2007}.
These findings have deeply affected our understanding of
self-organized networks and have been used for the investigation,
characterization and modeling of many different systems such as
infrastructure or biological and social networks.

Many works have studied the influence of complex network topologies
observed in real networks on the dynamical phenomena that unfold on
them \cite{Barrat:2008,Dorogovtsev:2008}. While these studies have
mostly focused on networks considered as static objects with a fixed
topology, networks' structures may in principle evolve, links may
appear and disappear. A first approach consists of assuming that links
are created and annihilated at a constant rate, independently of the
dynamical process that takes place on
them \cite{nota,Bianconi:2002,Bradde:2010}. Networks can however
display more interesting properties such as an adaptative behavior, in
which the dynamics on the network and of the network are related by feedback effects 
\cite{Bornholdt:2002,Marsili:2004,Holme:2006,MaxiSanMiguel:2008,Nardini:2008,Kozma:2008,Gross:2008}. 
Moreover, empirical investigations have shown that
link durations can significantly deviate from a Poisson
process \cite{Hui:2005,Scherrer:2008,Gautreau:2009,Sociopatterns,Cattuto:2010,Isella:2011}.

Social networks \cite{Granovetter:1973,Wasserman:1994} are prominent
examples of evolving networks. Social relationships are indeed
continuously changing, possibly in a way
correlated with the dynamical processes taking place during social
interactions (such as epidemic spreading or opinion
dynamics). Consequently, a number of works have been devoted to
modeling the dynamics of social interactions. Issues investigated in
this context are in particular community
formation \cite{Kumpula:2007,Johnson:2009,Palla:2007} and the evolution of
adaptive dynamics of opinions and social ties through schematic models
in which links can disappear or be rewired at random
\cite{Bornholdt:2002,Marsili:2004,Holme:2006,MaxiSanMiguel:2008,Nardini:2008,Kozma:2008,Gross:2008}.
Moreover, social networks evolve on many different timescales. The
static representation of social ties hide indeed dynamical sequences
of events such as face-to-face interactions, phone calls or email
exchanges, and can be measured by aggregating fast social interactions
over a certain period of time. 

Recently, technological advances have made possible the access to data
sets that give new insights into such link internal dynamics,
characterized by sequences of events of different durations.
Traces of human behavior are often unwittingly recorded in a
variety of contexts (financial transactions, phone calls, mobility
patterns, purchases using credit cards, etc.). Data have been gathered
and analyzed about the mobility patterns inside a
city~\cite{Chowell:2003}, between cities~\cite{Montis:2007}, as well
as at the country and at worldwide
levels \cite{Barrat:2004,Brockmann:2006,Gautreau:2009,Balcan:2009}.
At a more detailed level, mobile devices such as cell phones make it
possible to investigate individual mobility patterns and their
predictability~\cite{Gonzalez:2008,Song:2010}. Mobile devices and
wearable sensors using Bluetooth and Wifi technologies give access to
proximity patterns of pairs of
individuals \cite{Hui:2005,Eagle:2006,Kostakos,Pentland:2008,Salathe:2010},
and even face-to-face presence can be resolved with high spatial and
temporal resolution~\cite{Sociopatterns,Alani:2009,Cattuto:2010,Isella:2011}.
Finally, on-line interactions occurring between individuals can be
monitored by logging instant messaging or email
exchange~\cite{Eckmann:2004,Barabasi:2005,Kossinets:2006,Golder:2007,Leskovec:2008,Rybski:2009,Amaral:2009}.

The combination of these technological advances and of heterogeneous
data sources allows researchers to gather longitudinal data that have
been traditionally scarce in social network
analysis~\cite{Padgett:1993,Lubbers:2010}. Analysis of such data sets
has clearly shown the bursty nature of many human and social
activities, revealing the inadequacy of many traditional frameworks
that posit Poisson distributed processes. In particular, the
durations of ``contacts'' between individuals, as defined by the
proximity of these individuals, display broad distributions, as well
as the time intervals between successive
contacts \cite{Hui:2005,Scherrer:2008,Sociopatterns,Cattuto:2010,Salathe:2010,Isella:2011}.
Burstiness of interactions has strong consequences on dynamical
processes \cite{Holme:2005,Vazquez:2007,Onnela:2007,Havlin:2009,Latora:2009,Isella:2011},
and should therefore be correctly taken into account when modeling
the interaction networks. New frameworks are therefore
needed, which integrate the bursty character of human interactions and
behaviors into dynamic network models.

While a lot of modeling efforts have been devoted to static networks,
the development of models of dynamic networks has indeed until
recently attracted less attention
\cite{Gross:2008,Scherrer:2008,Hill:2009,Gautreau:2009,Stehle:2010}.
In a recent paper \cite{Stehle:2010}, we have presented an
agent-based modeling framework to describe how individuals interact
at short times scales in venues such as social gatherings (e.g.,
scientific conferences). The model is based on a reinforcement
dynamics (in the spirit of the preferential attachment in complex
networks \cite{Barabasi:1999} and of Hebbian learning) which might be
responsible for the bursty dynamics of social face-to-face
interactions. The proposed mechanism implies that {\em the longer an
agent is interacting in a group, the smaller is the probability that
he/she will leave the group; the longer an agent is isolated the smaller
is the probability that he/she will form a new group}.

In the present paper, we present an extensive characterization of the
model's properties, and show that it reproduces important features of
empirical data on social interactions at short timescales. We
characterize the rich phase diagram of the model, which includes
stationary and non-stationary regions. The analysis of the dynamical
properties of the model shows that it yields stationary broad
distribution of group lifetimes even if the underlying dynamics is
non-stationary, as also found in empirical data. In order to
illustrate the model's versatility, we give two examples of how it can
be extended to more realistic cases: in the first example, agents can
have heterogeneous propensities to form group with others; in the
second example, we introduce the possibility of a varying population,
where the number of individuals can be an arbitrary function, or
extracted from empirical data. In the two cases we show how properties
very close to the ones of real-world data sets can be obtained.
The proposed model is easily implementable, uses simple
but realistic mechanisms, reproduces a certain number of empirical facts,
and is amenable to further refinements. It can
therefore be used to produce artificial data sets of bursty interaction
networks on which dynamical phenomena can be simulated and studied.

The paper is organized as follows. In Sec. \ref{sec:data}, we review
the main properties of a representative empirical data set describing
the face-to-face proximity of individuals in social gatherings.
Section. \ref{sec:model} is devoted to the definition of the modeling
framework and to the analytical and numerical study of the simplest
versions of the model. Section. \ref{sec:extensions} is devoted to two
extensions of the model. We outline some conclusions and perspectives
in Sec. \ref{sec:conclusions}.

\section{Empirical data}
\label{sec:data}

The infrastructure developed by the SocioPatterns project, described
in \cite{Sociopatterns,Cattuto:2010,Isella:2011}, has yielded
measurements about the face-to-face proximity of individuals in
different types of social contexts (hospital, primary school,
scientific conferences, museum), with a fine grained time
resolution. The infrastructure is currently based on radio frequency
identification devices (RFID). Individuals participating to the data
collection are asked to wear small RFID tags on their chests (as a
conference badge), that emit radio packets at very low power. The
parameters of the infrastructure are tuned so that the tags can
exchange radio packets only when the individuals wearing them face
each other at close range (about 1 to 1.5 m), and so that face-to-face
proximity events (''contacts") can be assessed with very high accuracy
with a time resolution of 20s.  We present here for the sake of
completeness some characteristics of the data collected during the 6th
European Semantic Web Conference (ESWC, Heraklion, Greece) in 2009.
Analysis of other data sets can be found in Refs.
\cite{Sociopatterns,Cattuto:2010,Isella:2011}, with very similar
features.

\begin{figure}[h]
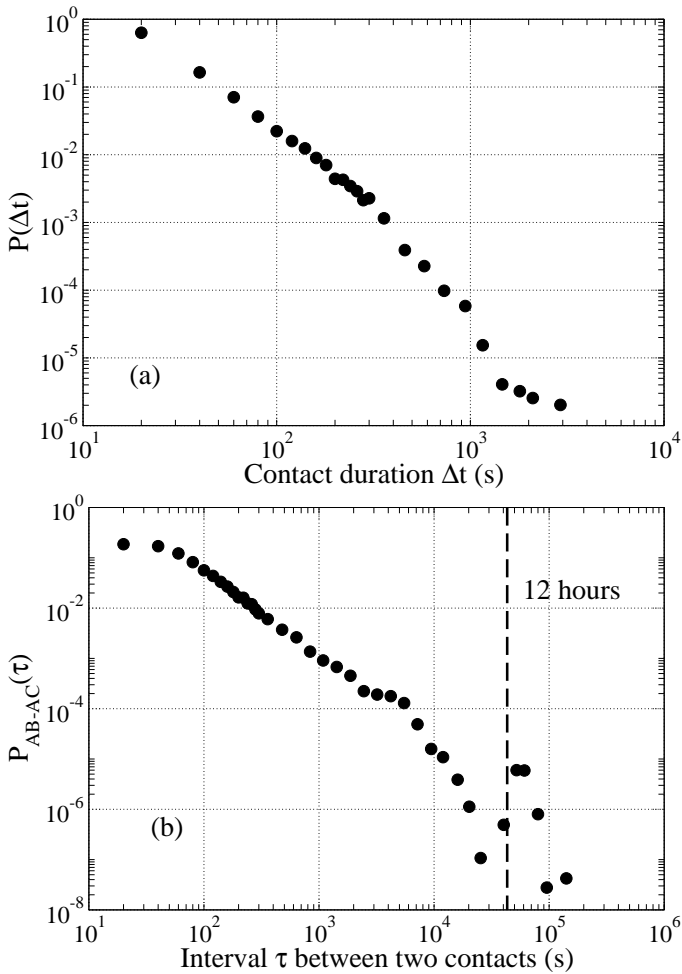

\includegraphics[width=0.5\textwidth]{fig1a}
\includegraphics[width=0.5\textwidth]{fig1b}
\caption{\label{fig:Distribution-of-contact}Distribution of contact
  durations (a) and of time intervals between two contacts involving a
  common individual (b).}
\end{figure}

\begin{figure}
\includegraphics[width=0.5\textwidth]{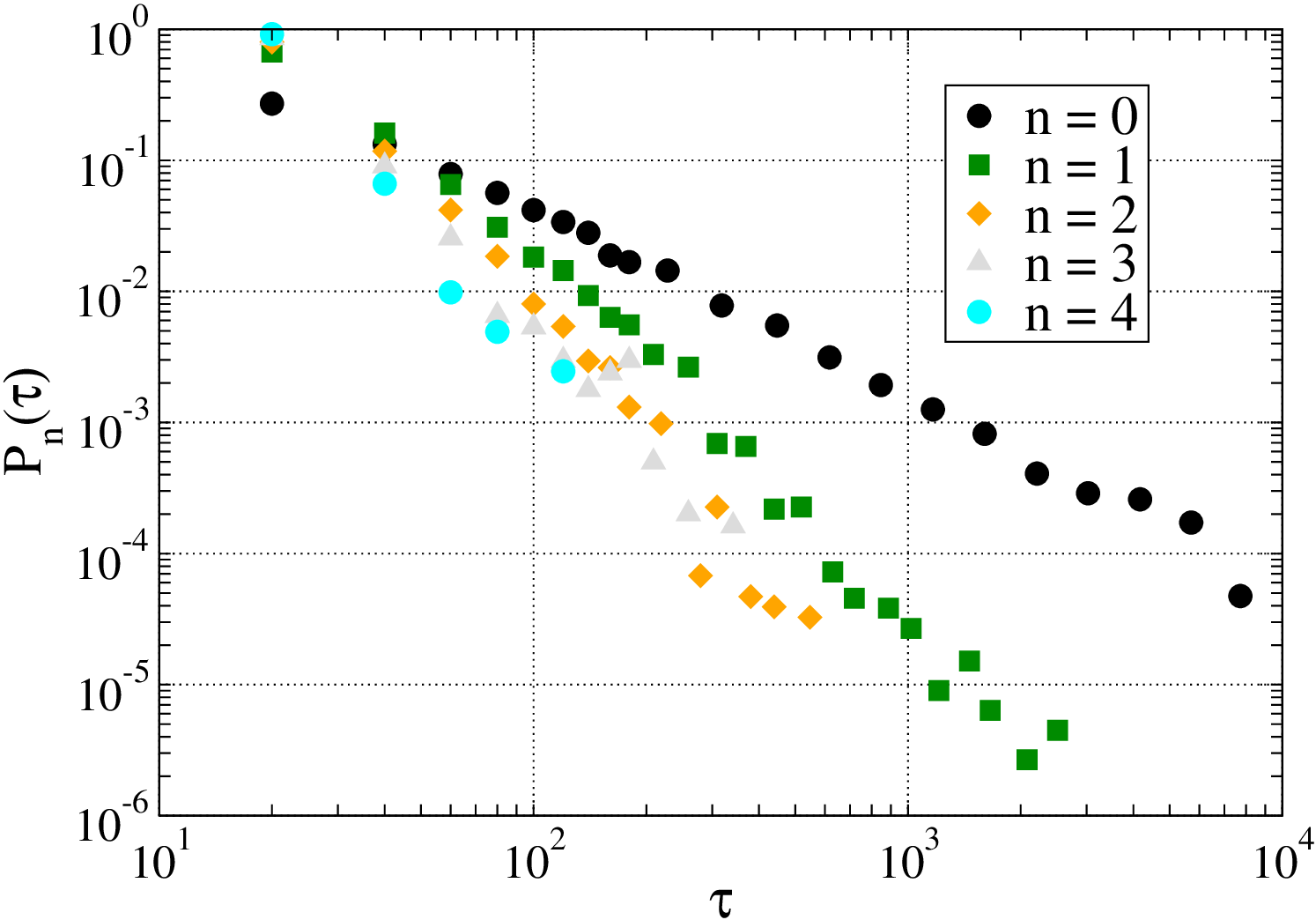}
\caption{(Color online) Distributions $P_n(\tau)$ of the durations (in
  seconds) of groups of size $n+1$ for the ESWC data set.}
\label{fig:state_duration_eswc}
\end{figure}

Face-to-face proximity patterns have been collected for $175$
voluntary participants (among the $305$ conference attendants) over
$3$ days.  $14520$ contact events have been registered, corresponding
to $27.3$ contacts per individual per day. The contact durations
display a very broad distribution, close to a power law, (shown in
Fig. \ref{fig:Distribution-of-contact}): the average duration is of
$46$ s, but long-lasting contacts are as well observed, and no
characteristic contact timescale can be extracted from the data.
Figure \ref{fig:Distribution-of-contact} also displays a quantity of
interest, namely the distribution of time intervals between the start
of two consecutive contacts of a given individual $A$ with two
distinct persons $B$ and $C$. In other words, if $A$ starts a contact
with $B$ at time $t_{AB}$ , and then starts a different contact with
$C$ at $t_{AC}$, the inter-contact interval is defined as $\tau=t_{AC}
- t_{AB}$.  These time intervals constrain causal processes such as
information diffusion or epidemic spreading, as they determine the
timescale after which an individual receiving some information or
disease is able to propagate it to another individual. As shown in
Fig. \ref{fig:Distribution-of-contact}, broad distributions are also
found in this case. We also show in
Fig. \ref{fig:state_duration_eswc} the distributions of lifetimes of
groups of size $n+1$ ($n=0$ corresponds to an isolated person, $n=1$
to a pair, etc). All these distributions are broad, compatible with
power-law shapes, and become narrower for increasing $n$ (larger
groups are less stable than smaller ones).

\begin{figure}[tp]
\includegraphics[width=0.5\textwidth]{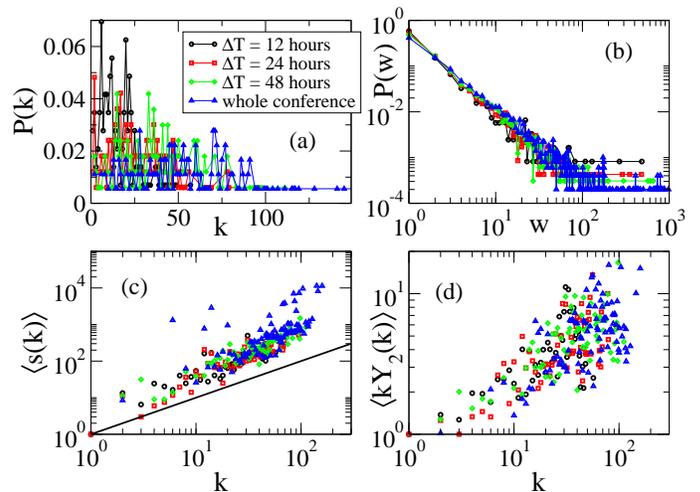}
\caption{(Color online) Aggregated networks' characteristics for
the empirical ESWC data: (a) degree
distribution, (b) weight distribution, (c) average strength of nodes of degree
$k$, vs $k$, and (d) average Herfindahl-Hirschman index of nodes of
degree $k$, vs $k$.}
\label{fig:aggr_nets_eswc}
\end{figure}

The burstiness of the contact pattern revealed by the broad
distribution of contact durations has consequences also on aggregated
views of the dynamical contact network. Aggregated contact networks
over a given time window are defined as follows: each node corresponds
to an individual, and an edge is drawn between two nodes if at least
one contact event has been registered during the time window between
the two corresponding individuals. Each edge is weighted by the sum of
the contact durations between these individuals during this time
window.  As shown in Fig. \ref{fig:aggr_nets_eswc} (top right), the
distributions of such weights are broad, independent of the time
window considered (see also \cite{Cattuto:2010,Isella:2011}).

Figure \ref{fig:aggr_nets_eswc} also displays other characteristics of
aggregated networks constructed with time windows of different
lengths. As also found in \cite{Cattuto:2010,Isella:2011} for other
deployments of the same infrastructure, the distribution of degrees
(the degree gives the number of distinct individuals with whom a given
individual has been in contact) are not broadly distributed. This
behavior is in contrast with the degree distribution of many
empirically studied social networks \cite{Liljeros:2001,Newman:2001,Barrat:2008}.
It should however be emphasized that we are here dealing with
face-to-face interactions occurring in a restricted environment among
a relatively small population and on short timescales, while studies
such as \cite{Liljeros:2001,Newman:2001} are concerned with social ties
created and defined over much longer times cales. 
Moreover, we note that a narrow degree distribution (a power law with a large exponent) has also 
been found in the degree distribution of networks defined by phone-calls data sets \cite{Onnela:2007}. 

The strength $s_i$ of a node $i$ is defined as the sum of the
weights $w_{ij}$ between $i$ and its neighbors $j$, that is,
\begin{equation}
s_i=\sum_j w_{ij}.
\end{equation}
It gives the cumulated durations of the interactions of the corresponding
individual. The average strength of nodes of degree $k$ [$\langle
s(k)\rangle$] indicates how the weights are distributed.
If the weights are uniformly distributed among the links of the networks,
the average strength $\avg{s(k)}$ grows linearly with $k$, that is,
$\avg{s(k)}\propto k\avg{w}$.
On the contrary, if stronger ties are more frequently linked to 
highly connected agents, a superlinear behavior of $\avg{s(k)}$ versus $k$ is observed \cite{Barrat:2004}.
The RFID data on face-to face interaction are consistent with a linear or slightly
faster behavior of $\avg{s(k)}$ versus $k$  \cite{Cattuto:2010,Isella:2011}, hinting at a weak
correlation between weights and degrees. 
Correlations between network topology and distribution of the weights have been found in
various complex networks with broad degree distribution \cite{Barrat:2004},
and several models based on reinforcement dynamics have been proposed in this context
\cite{Barrat:2004b,Bianconi:2005,Wang:2005}.

Another measure of the weights' distribution is given by the
Herfindahl-Hirschman index $Y_2$
\cite{hirschman:1964,herfindahl:1959}, also known in the physics
literature under the name of ''participation ratio".  This index,
defined as
\begin{equation}
Y_2(i)= \sum_{j \in \mathcal{N}(i)} \left( \frac{w_{ij}}{s_i} \right)^2 ,
\end{equation}
where $\mathcal{N}(i)$ refers to the set of neighbors of $i$, gives a
measure of the heterogeneity of the weights among the neighbors of a
node. When all weights $w_{ij}$ of the links connected to node $i$ are
equal, that is, $w_{ij}=s_i/k_i$, this index is inversely proportional to
the degree $k_i$ of node $i$, that is,
\begin{equation}
Y_2(i)=\frac{1}{k_i} .
\end{equation}
On the contrary, when one of the links has a much larger weight than
the others, $Y_2(i)$ is close to $1$.  The departure of $k_i Y_2(i)$
from $1$ thus indicates the local heterogeneity of weights around each
node. Figure \ref{fig:aggr_nets_eswc} shows that the average of this
quantity over nodes of degree $k$ [$k Y_2(k)$], is
larger than $1$ and increases with $k$: each individual divides
his/her time unevenly among her/his contacts, and this behavior
becomes more pronounced as the number of distinct contacted persons
increases.

\section{A model for social dynamics at short timescales}
\label{sec:model}

In this section, we present a model describing the social dynamics of
$N$ agents forming small groups of different size. In this model,
each agent can interact with any other agent. This model describes
therefore social interactions within a closed environment of
relatively small size where agents are free to meet, such as a
conference venue. We assign to each agent $i=1,2,\ldots, N$ a
coordination number $n_i=0,1,2\ldots, N$ indicating the number of
agents interacting with him/her. If an agent $i$
has coordination number $n_i=0$ he/she is isolated, and if 
$n_i=n>0$ he/she is part of a group of $n+1$ agents, who all interact
with each other (thus forming a clique).
We also assign to each agent $i$ the temporal variable $t_i$ indicating
the last time at which his/her coordination number $n_i$ has changed.
 
The dynamics of the model is as follows. Starting from random initial
conditions, at each time step $t$ the following steps are performed:
\begin{itemize}
\item[(1)] An agent $i$ is chosen randomly.
\item[(2)] The agent $i$ updates his/her coordination number $n_i=n$
  with a certain probability $p_n(t,t_i)$ that may depend on the
  agent's state, on the present time $t$, and on the last time $t_i$ at
  which $i$'s state evolved. With probability $1-p_n(t,t_i)$, the
  agent does not change state.

If the coordination number $n_i$ is updated, the action of the agent is chosen with the following rules.
\begin{itemize}
\item[(i)] If the agent $i$ is isolated, that is, $n_i=0$, he/she starts an
  interaction with another isolated agent $j$ chosen with probability
  proportional to $p_0(t, t_j)$. The coordination number of the agent
  $i$ and of the agent $j$ are then updated according to the rule $n_i
  \rightarrow 1$ and $n_j \rightarrow 1$.
\item[(ii)] If the agent $i$ is interacting in a group,
  that is, $n_i=n>0$, with probability $\lambda$ the agent leaves the
  group and with probability $(1-\lambda)$ he/she introduces an isolated
  agent in the group. If the agent $i$ leaves the group, his/her
  coordination number is updated ($n_i \rightarrow 0$), as well as the
  coordination numbers of all the agents in the original group,
  that is, $n_k \rightarrow n-1$ (for all agents $k$ in
  the original group). On the contrary, if the agent $i$ introduces
  another isolated agent $j$ to the group, $j$ is chosen
  with probability proportional to $p_0(t,t_j)$ and the coordination
  numbers of all the interacting agents are changed according to the
  rules  $n_j \rightarrow n+1$ and $n_k \rightarrow n+1$ (for all $k$ in the group).
\end{itemize}
\end{itemize}

The structure and properties of the interactions between agents depend
on the choice of the probabilities $p_n$, which control the tendency
of the agents to change their state, and on the parameter $\lambda$,
which determines the tendency either to leave groups or on the
contrary to make them grow. The simplest choice consists of
considering constant probabilities $p_n(t,t')=p_n$: at each time,
every agent has a fixed probability to form a group or split from a
group. In this case the formation of the groups is a Poisson process,
and the distributions of the durations of contacts between agents, 
or  the lifetime of a group, are exponentially distributed.

As recalled in the introduction however, the distributions of the
times describing human activities are typically broad
\cite{Barabasi:2005,Vazquez:2007,Hui:2005,Kostakos,Rybski:2009,Cattuto:2010,Isella:2011},
and are clearly closer to power-laws that lack a characteristic time
scale than to exponentials. A possible explanation of such results is
given by mechanisms in which the decisions of the agents to form or
leave a group are driven by memory effects dictated by reinforcement
dynamics, that can be summarized in the following statement: {\em the
  longer an agent is interacting in a group, the smaller is the
  probability that he/she will leave the group; the longer an agent is
  isolated, the smaller is the probability that he/she will form a new
  group}. In particular, such reinforcement principle implies that the
probabilities $p_n(t,t')$ that an agent with coordination number $n$
changes his/her state depend on the time elapsed since his/her last
change of state. A simple way to introduce this hypothesis is to
consider functions $p_n(t,t')=p_n(t-t')$. Reinforcement mechanisms are
then described by decreasing functions $p_n$. We will see in the next
subsections how the evolution equations of the number of agents in
each state can be written at the mean-field level for arbitrary
functions $p_n$. Finding solutions of this set of equations is however
not always possible, and we will focus on functions $p_n$ scaling as
$1/(t-t')$, for two reasons: on the one hand, it represents one of the
cases that is fully amenable to analytical computations and, on the
other hand, such a scaling behavior is needed to obtain power-law
distributions for the contact durations and thus dynamical
characteristics compatible with empirical data.
We consider in particular functions given by
\begin{equation}
p_n(t,t')=\frac{b_n}{1+(t-t')/N} , 
\label{p}
\end{equation}
and, in order to reduce the number of parameters, we
moreover take $b_n=b_1$ for every $n\geq1$, indicating the fact that
interacting agents change their state independently on the number of
other agents $n$ with whom they are interacting, provided that $n
\geq 1$. The model's parameter are thus $b_0$, $b_1$, and $\lambda$.

\subsection{Pairwise interactions}

Let us first consider a restricted version of the model, in which the
agents can only interact in pairs. This set-up is obtained by setting
$\lambda=1$ and by considering initial conditions in which the agents
interact at most in groups of size $2$. In this case, each agent is
thus assigned a variable $n_i=0,1$ indicating if the agent $i$ is
isolated $(n_i=0)$ or interacting with another agent $(n_i=1)$. 

As in the analysis of empirical data, the most immediate quantities of
interest concern the time spent by agents in each state, the duration
of contacts between two agents, and the time intervals between
successive contacts of an agent. To gain insight into these temporal
properties of the system, we can write rate equations for the
evolution of the numbers $N_n(t,t')$ of agents in state $n=0$ at time
$t$ who have not changed state since time $t'$. In the mean-field
approximation, and treating time and numbers as continuous variables,
these equations are given by 
\bea 
\frac{\partial N_0(t,t')}{\partial
  t}&=&-2\frac{N_0(t,t')}{N}p_0(t,t')+\pi_{10}(t)\delta_{tt'},\nonumber \\ 
\frac{\partial N_1(t,t')}{\partial t}&=&-2\frac{N_1(t,t')}{N}p_1(t,t')+\pi_{01}(t)\delta_{tt'},
   {\label{dN01}} 
\eea 
where the transition rates $\pi_{n,m}(t)$ denote the average number of
agents switching their states from $n$ to $m$ ($n \rightarrow m$) at
time $t$. If the agents make their decisions according to the
reinforcement dynamics described by the probabilities $p_n(t,t')$
given by Eq. (\ref{p}), the dynamic equations (\ref{dN01}) have a
solution of the form
\bea
N_0(t,t')&=&\pi_{10}(t')\left(1+\frac{t-t'}{N}\right)^{-2b_0} \,,
\nonumber\\ 
N_1(t,t')&=&\pi_{01}(t')\left(1+\frac{t-t'}{N}\right)^{-2b_1} \,.
\label{N1}
\eea
Since the total number of isolated agents who 
change their state at time $t$ is equal to $\pi_{01}(t)$ and the total number of interacting agents 
who change their state is equal to $\pi_{10}(t)$, it follows that  
$\pi_{10}(t)$ and $\pi_{01}(t)$ are given in terms of $N_0(t,t')$ and $N_1(t,t')$ by the relations
\bea
\pi_{10}(t)&=&\frac{2}{N}\sum_{t'=1}^tp_1(t,t')N_1(t,t'),\nonumber \\
\pi_{01}(t)&=&\frac{2}{N}\sum_{t'=1}^tp_0(t,t')N_0(t,t') .
\label{pi01}
\eea
To solve the coupled set of equations (\ref{N1}) and (\ref{pi01}), 
we assume self-consistently that $\pi_{10}(t)$ and $\pi_{01}(t)$ are either 
constant or decaying in time as power laws. Therefore, we assume
\bea
\pi_{10}(t)&=&\tilde{\pi}_{10}\left(\frac{t}{N}\right)^{-\alpha_0},\nonumber\\
\pi_{01}(t)&=&\tilde{\pi}_{01}\left(\frac{t}{N}\right)^{-\alpha_1} .
\label{pi01f}
\eea

\begin{figure}
\includegraphics[width=0.5\textwidth]{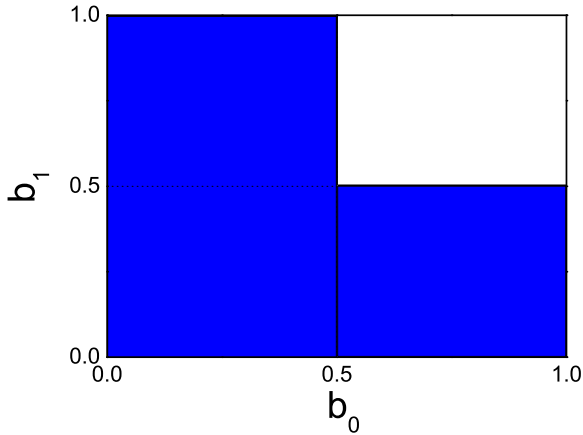}
\caption{(Color online) Phase diagram of the pairwise model. The white
  area indicates the stationary regime in which the transition rate is
  constant. The colored (gray) area indicates the non-stationary
  phase.}
\label{Fig2}
\end{figure}

\begin{figure}
\includegraphics[width=0.5\textwidth]{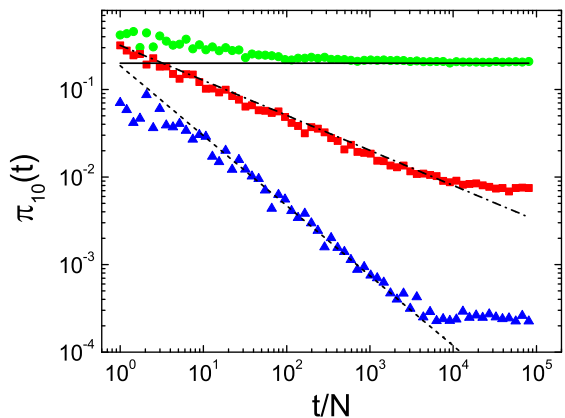}
\caption{(Color online) Evolution of the transition rate $\pi_{10}(t)$ in the
  different phase regions of the pairwise interaction model. The
  simulation is performed with $N=1000$ agents for a number of time
  steps $T_{max}=N\times 10^5$, and averaged over $10$
  realizations. The simulations are performed in the stationary region
  with parameter values $b_0=b_1=0.7$ (circles) and in the
  non-stationary region with parameter values $b_0=0.3$, $b_1=0.7$
  (squares) and $b_0=b_1=0.1$ (triangles). The lines indicate the
  analytical predictions Eqs. (\ref{pi01f})-(\ref{pitilde}).}
\label{Fig1}
\end{figure}

To check the self-consistent assumption Eq. (\ref{pi01f}), we insert
it in Eqs. (\ref{N1}) and (\ref{pi01}) and compute the values of the
parameters $\alpha_0$, $\alpha_1$, $\tilde{\pi}_{10}$ and $\tilde{\pi}_{01}$ 
that determine the solution in the asymptotic limit $t\to \infty$. If
$\alpha_0=\alpha_1=0$, we obtain a stationary solution in which
$\pi_{10}(t)=\tilde{\pi}_{10}$ and $\pi_{01}(t)=\tilde{\pi}_{01}$, are
independent of time. On the contrary if $\alpha_0>0$ or $\alpha_1>0$, the
system is non-stationary, with transition rates $\pi_{10}(t)$ and
$\pi_{01}(t)$ decaying in time. The system dynamics slows down. In
Appendix \ref{AA}, we give the details of this self-consistent
calculation in the large $N$ limit, which yields
$\alpha_0=\alpha_1=\alpha$ and
$\tilde{\pi}_{10}=\tilde{\pi}_{01}=\tilde{\pi}$, with 
\bea \alpha &=&\max{(0,1-2b_1,1-2b_0)}\ ,\nonumber \\ 
\tilde{\pi}&=&\frac{\sin{[2\pi
      \min{(b0,b1)}]}}{\pi}[1-\delta(\alpha,0)] \nonumber \\ 
&&+\frac{(2b_0-1)(2b_1-1)}{2(b_0+b_1-1)}\delta(\alpha,0).
\label{pitilde}
\eea 

\begin{figure}
\includegraphics[width=0.5\textwidth]{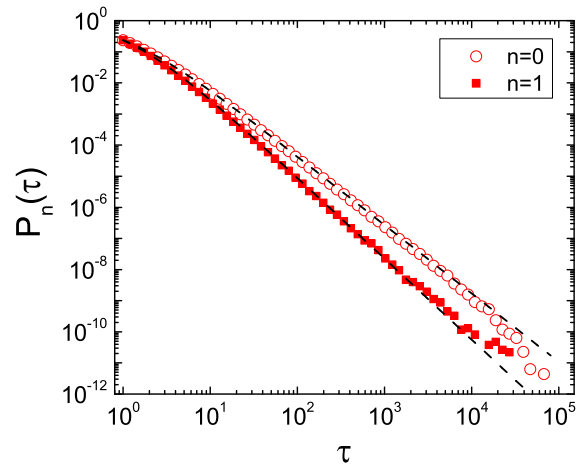}
\caption{(Color online) Probability distribution of the durations of
  contacts $P_1(\tau)$ and of the inter-contact durations $P_0(\tau)$ in the
  stationary region, for the pairwise model. The data is reported for
  a simulation with $N=1000$ agents, run for $T_{max}=N\times 10^5$
  elementary time steps, with parameter values $b_0=0.6$, $b_1=0.8$. The
  data is averaged over $10$ realizations. }
\label{FigP}
\end{figure}

\begin{figure}
\includegraphics[width=0.5\textwidth]{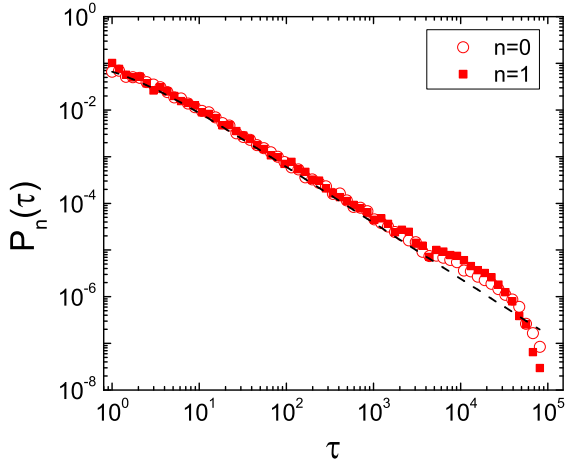}
\caption{(Color online) Probability distribution of the durations of
  contacts $P_1(\tau)$ and of the inter-contact durations $P_0(\tau)$ in the
  non-stationary region of the pairwise model, with $b_0<0.5$ and
  $b_1<0.5$. In this region we observe some deviations of the
  probabilities $P_1(\tau)$ and $P_{0}(\tau)$ from the power-law
  behavior for large durations. The data are reported for a simulation
  with $N=1000$ agents run for $T_{max}=N\times 10^5$ elementary time
  steps, with parameter values $b_0=b_1=0.1$. The data are averaged
  over $10$ realizations.}
\label{condensation}
\end{figure}

The analytically predicted dynamical behavior or the model can be
summarized by the phase diagram depicted in Fig. \ref{Fig2} (that we
discuss now in more detail), together with the numerical simulations of
the stochastic model displayed in Fig. \ref{Fig1}.
\begin{itemize}
\item{\em Stationary region ($b_0>0.5$ and $b_1>0.5$) -} In this
  region of the phase diagram, the self-consistent equation predicts
  $\alpha=0$, so that a stationary state solution is expected, where
  $\pi_{01}(t)=\tilde{\pi}$ is given by Eq. (\ref{pitilde}). In this
  stationary state the number of isolated agents and the number or
  interacting agents are constant on average, but the dynamics is not
  frozen, since $\tilde{\pi} > 0$: agents continuously form and leave
  pairs. The simulations shown in Fig. \ref{Fig1} for $b_0=b_1=0.7$
  confirm this analytical prediction.

 \item {\em Non stationary region ($b_0<0.5$ or $b_1<0.5$) -} 
   In this region of the
   phase diagram, the self-consistent equation predicts a
   non-stationary solution with $\pi_{10}(t)$ and $\pi_{01}(t)$
   decaying with $t$ as a power-law of exponent
   $\alpha=\max(1-2b_0,1-2b_1)$.  Figure \ref{Fig1} shows such a decay
   for $b_0=0.3$, $b_1=0.7$ and for $b_0=b_1=0.1$, which is however
   truncated by finite-size effects for $t$ larger than
   $t_c(N)\propto N$. Therefore the system eventually becomes
   stationary with a very slow dynamics (very small transition rates
   $\pi_{10}(t)$ and $\pi_{01}(t)$).
 \end{itemize}

Empirical studies often focus on the statistics of contact durations
between individuals, and of the time intervals between two contacts of
a given individual. These quantities of interest can be computed in
our model, respectively, as the probabilities ${P}_{1}(\tau)$ that an
agent remains in a pair during a time $\tau=(t-t')/N$, and
${P}_{0}(\tau)$ that an agent remains isolated for a time interval
$\tau=(t-t')/N$. These probabilities are determined by the numbers of
agents in each state and the rates at which the agents change their
state. The probability distributions of the time spent in each state, 
integrated between the initial time and an arbitrary time $t$, are given by
\bea
P_n(\tau)\propto \int_{t'=0}^{t-N\tau} p_{n}(t'+N\tau,t')N_{n}(t'+N\tau,t') dt'
\label{P}
\eea 
for $n=0,1$. Inserting the expression given by Eq. (\ref{N1})
for $N_n(t,t')$ and the definition of $p_n(t,t')$ given by
Eq. (\ref{p}) in Eq. (\ref{P}), we obtain the power-law distributions 
\bea
P_n(\tau)\propto\left(1+\tau \right)^{-2b_n-1} 
\label{Pn}
\eea 
for $n=0,1$. These analytical predictions are compared with
numerical simulations in Fig. \ref{FigP} for $b_0=0.6$, $b_1=0.8$
(stationary system) and in Fig. \ref{condensation} for $b_0=b_1=0.1$
(non-stationary $\pi_{10}$ and $\pi_{01}$). Interestingly, even when
the system is non-stationary, the distributions $P_n(\tau)$ remain
stationary.

\subsection{Formation of groups of any size}

In this subsection we extend the solution obtained for the pairwise
model to the general model with arbitrary value of the parameter
$\lambda$, where groups of any size can be formed. Therefore the
coordination number $n_i$ of each agent $i$ can take any value up to
$N-1$. Extending the formalism used in
the previous subsection, we denote by $N_n(t,t')$ the number of
agents with coordination number $n=0,1,\ldots,N-1$ at time $t$, who
have not changed state since time $t'$. In the mean field
approximation, the evolution equations for $N_n(t,t')$ are given by
  
\bea
\frac{\partial N_0(t,t')}{\partial t}&=&-2\frac{N_0(t,t')}{N}p_0(t,t')-(1-\lambda)\epsilon(t)\nonumber \\
&&\times \frac{N_0(t,t')}{N}p_0(t,t')+\sum_{i \ge 1}\pi_{i,0}(t)\delta_{tt'},\nonumber\\
\frac{\partial N_1(t,t')}{\partial t}&=&-2\frac{N_1(t,t')}{N}p_1(t,t') \nonumber \\
&&+[\pi_{0,1}(t)+\pi_{2,1}(t)]\delta_{tt'}, \nonumber \\
\frac{\partial N_n(t,t')}{\partial t}&=&-(n+1) \frac{N_n(t,t')}{N}p_1(t,t')  \nonumber \\
&&\hspace*{-20mm}+[\pi_{n-1,i}(t)+\pi_{n+1,n}(t)+\pi_{0,n}(t)]\delta_{tt'},~~n\ge 2. 
\label{dNiB}
\eea

In these equations, the parameter $\epsilon(t)$ indicates the rate
at which isolated nodes are introduced by another agent in already
existing groups of interacting agents. Moreover, $\pi_{mn}(t)$
indicates the transition rate at which agents change coordination
number from $m$ to $n$ (i.e. $m\to n$) at time $t$. In the mean-field
approximation the value of $\epsilon(t)$ can be expressed in terms of
$N_n(t,t')$ as
\be \epsilon(t)=\frac{\sum_{n \ge
    1}\sum_{t'=1}^tN_n(t,t')p_1(t,t')}{\sum_{t'=1}^tN_0(t,t')p_0(t,t')}.
\label{epsilon}
\ee
In the case of reinforcement dynamics described by the probabilities $p_n(t,t')$
given by Eq. (\ref{p}), and 
assuming that asymptotically in time $\epsilon(t)$ converges to a time-independent variable, that is, 
$\lim_{t\to\infty}\epsilon(t)=\hat{\epsilon}$, the solution to the rate
equations (\ref{dNiB}) in the large time limit is given by
\bea
\label{NiB}
N_0(t,t')&=&N_0(t',t')\bigg(1+\frac{t-t'}{N}\bigg)^{-b_0[2+(1-\lambda)\hat{\epsilon}]}, \nonumber\\
N_1(t,t')&=&N_1(t',t')\bigg(1+\frac{t-t'}{N}\bigg)^{-2b_1}, \\
N_n(t,t')&=&N_n(t',t')\bigg(1+\frac{t-t'}{N}\bigg)^{-(n+1)b_1}  ~~\ \mbox{for} \ n\geq 2,\nonumber
\eea
with
\bea
N_0(t',t')&=&\sum_{n \ge 1}\pi_{n,0}(t'),\nonumber \\
N_1(t',t')&=&\pi_{0,1}(t')+\pi_{2,1}(t'),  \\
N_n(t',t')&=&\pi_{n-1,n}(t')+\pi_{n+1,n}(t')+\pi_{0,n}(t')~~ \ \mbox{for} \ n\geq 2.\nonumber
\label{pig}
\eea

The transition rates ${\pi_{m,n}(t)}$ can be determined in terms of
$N_n(t,t')$ as shown in the Appendix \ref{A2}. In
order to solve the equations we make the further assumption that the
transition rates $\pi_{mn}(t)$ are either constant or decaying with
time according to a power law, that is. 
\be
\pi_{m,n}(t)=\tilde{\pi}_{m,n}\left(\frac{t}{N}\right)^{-\alpha_{m,n}} \ .
\label{pipq}
\ee
Self-consistent calculations (see Appendix \ref{A2}) 
determine the value of the quantities
$\hat{\epsilon}$, $\alpha_{mn}$, and $\tilde{\pi}_{mn}$. For $\lambda>0.5$ the
self-consistent assumption Eq. (\ref{pipq}) is valid and we find, as
in the case of pairwise interactions, that $\alpha_{m,n}=\alpha$
$\forall (m,n)$, with
\be 
\alpha=\max\left(0, 1-b_0\frac{3\lambda-1}{2\lambda-1}, 1-2b_1\right) .
\label{ag}
\ee
This solution generalizes the case of the pairwise model, which is
recovered by setting $\lambda=1$. For $\lambda\leq 0.5$ 
the self-consistent assumption breaks down and we will 
resort to numerical simulations.

The probability distributions of the time spent in each state, 
integrated between the initial time and an arbitrary time $t$, are given by
\bea
P_n(\tau)\propto \int_{t'=0}^{t-N\tau} p_{n}(t'+N\tau,t')N_{n}(t'+N\tau,t') dt'.
\label{P2}
\eea 
Inserting the expression given by Eq. (\ref{NiB})
for $N_n(t,t')$ and the definition of $p_n(t,t')$ given by
Eq. (\ref{p}) in Eq. (\ref{P2}), we obtain the power-law distributions 
\bea \nonumber
P_0(\tau)&\propto &\left(1+\tau \right)^{-b_0[2+(1-\lambda \hat{\epsilon})]-1} \\
P_n(\tau) &\propto &\left(1+\tau \right)^{-(n+1)b_1-1} \ \mbox{for} \ n\geq 1. \label{Pn2}
\eea

\begin{figure}
\includegraphics[width=0.5\textwidth]{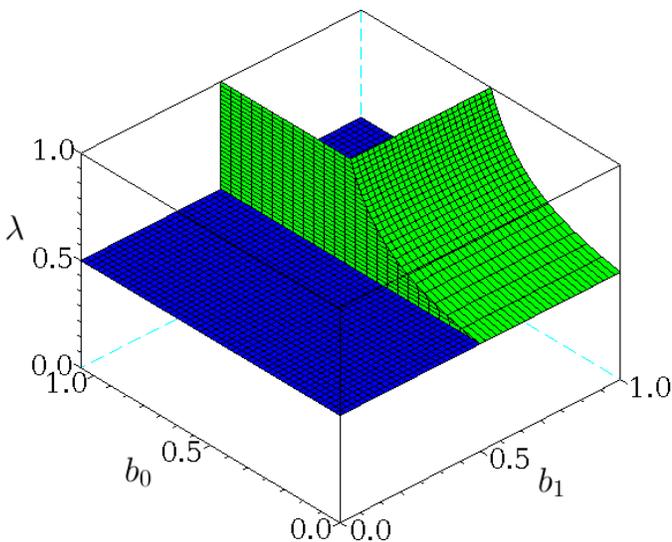}
\caption{(Color online) Phase diagram of the general model with formation of groups
  of arbitrary size. The region behind the green surface corresponds
  to the stationary phase [i.e., region (I), with $\lambda>0.5$,
  $b_1>0.5$ and $b_0>\frac{2\lambda -1}{3 \lambda-1}$]. The region in
  front of the green surface and above the blue one [region (II)]
  corresponds to a non-stationary system with decaying transition
  rates.  Strong finite size effects with a temporary formation of a
  large cluster are observed in the region below the blue surface
  [i.e., region (III) with $\lambda < 0.5$].}
\label{Fig3}
\end{figure}

\begin{figure}
\includegraphics[width=0.5\textwidth]{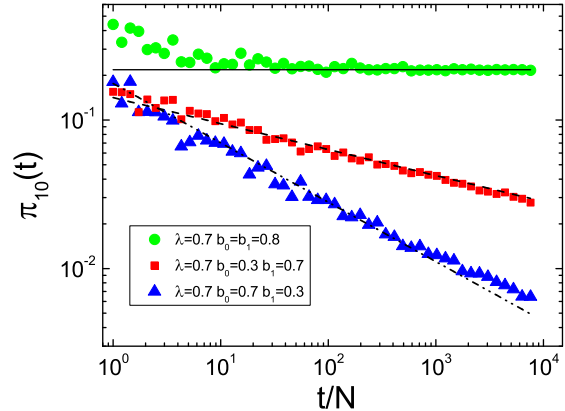}
\caption{(Color online) Transition rate $\pi_{10}(t)$ for the model in
  the presence of groups of any size, for different parameters $\lambda$,
  $b_0$, $b_1$ corresponding to the different regions of the phase
  diagram. The straight lines correspond to the analytical
  predictions.  The simulation is performed with $N=1000$ agents for a
  number of time steps $T_{max}=N\times 10^4$.  The data are averaged
  over $10$ realizations. }
\label{p10m}
\end{figure}

\begin{figure}
\includegraphics[width=0.5\textwidth]{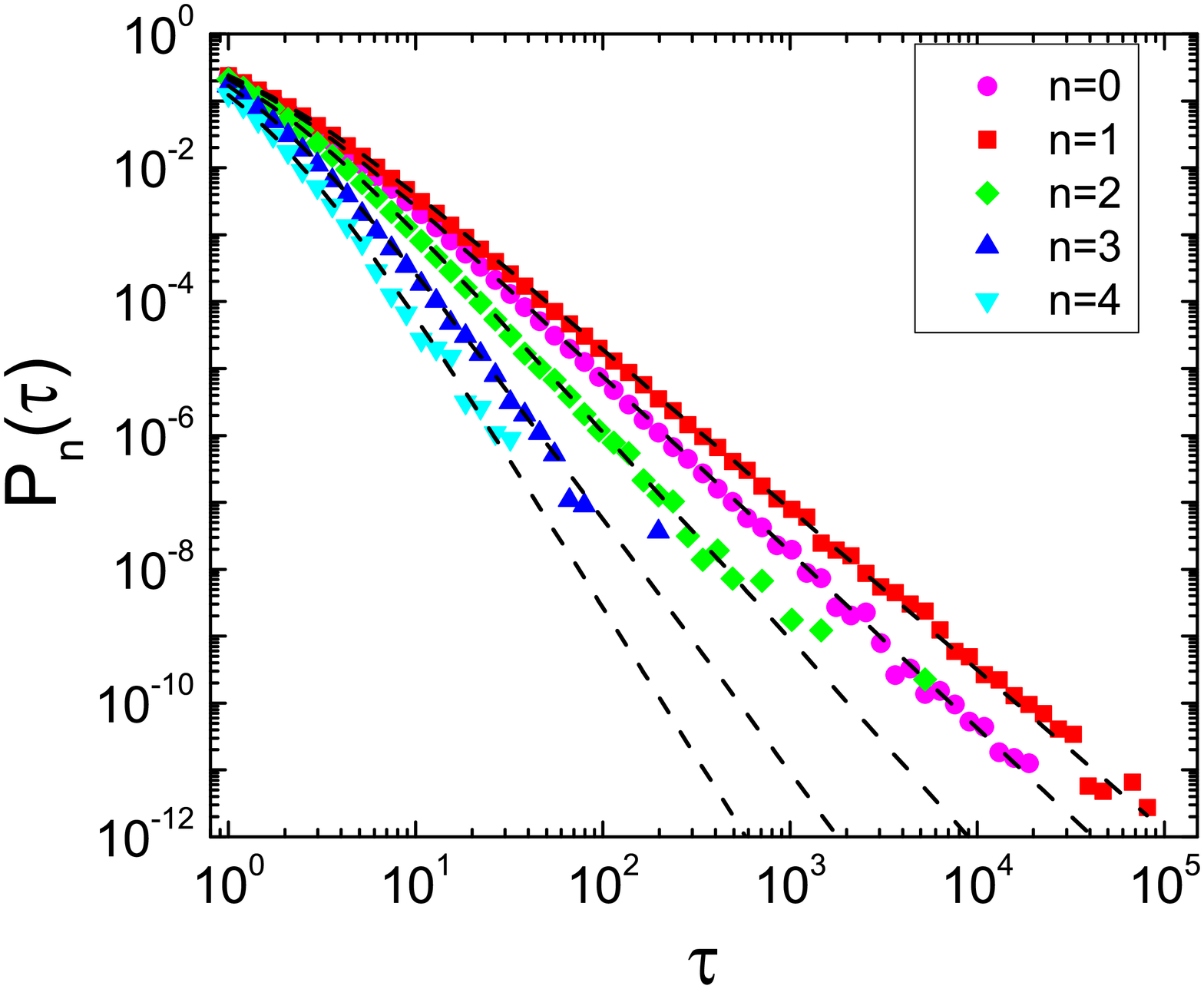}
\caption{(Color online) Distribution $P_n(\tau) $ of durations of
  groups of size $n+1$ in the stationary region. The simulation is
  performed with $N=1000$ agents for a number of time steps
  $T_{max}=N\times 10^5$. The parameter used are $b_0=b_1=0.7$,
  $\lambda=0.8$. The data are averaged over $10$ realizations. The
  dashed lines correspond to the analytical predictions
  Eqs. (\ref{Pn2}).}
\label{Groups_stationary}
\end{figure}

\begin{figure}
\includegraphics[width=0.5\textwidth]{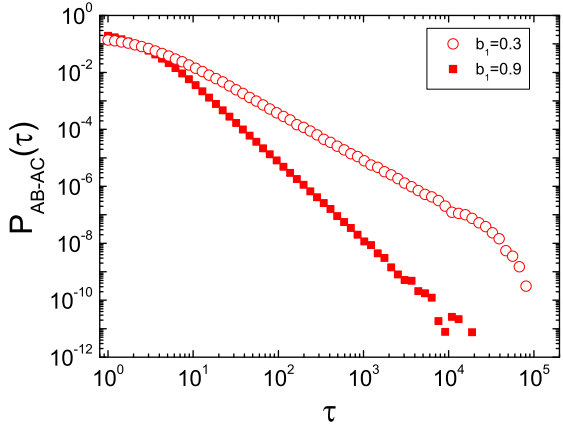}
\caption{(Color online) Distribution of time intervals between
  successive contacts of an individual, for 
  $\lambda=0.8$, $b_0=0.7$ and $b_1=0.3$ and $0.9$. The
  simulation is performed with $N=10^4$ for a number of time steps
  $T_{max}=N\times 10^5$. The data are averaged over $10$
  realizations.}
\label{Pabac}
\end{figure}

\begin{figure}
\includegraphics[width=0.5\textwidth]{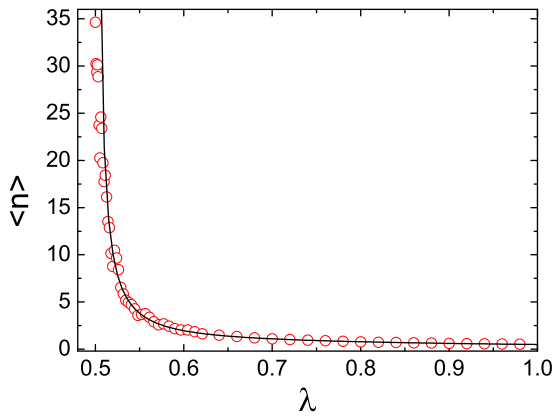}
\caption{(Color online) Average coordination number $\avg{n}$ vs
  $\lambda$ for $b_0=b_1=0.7$ . The simulation is performed with
  $N=2000$ agents for a number of time steps $T_{max}=N \times
  10^3$. $\avg{n}$ is computed in the final state over $30$
  realizations. The solid line indicates the theoretical prediction
  given by Eq. $(\ref{an})$.}
\label{avgn}
\end{figure}

\begin{figure}
\includegraphics[width=0.5\textwidth]{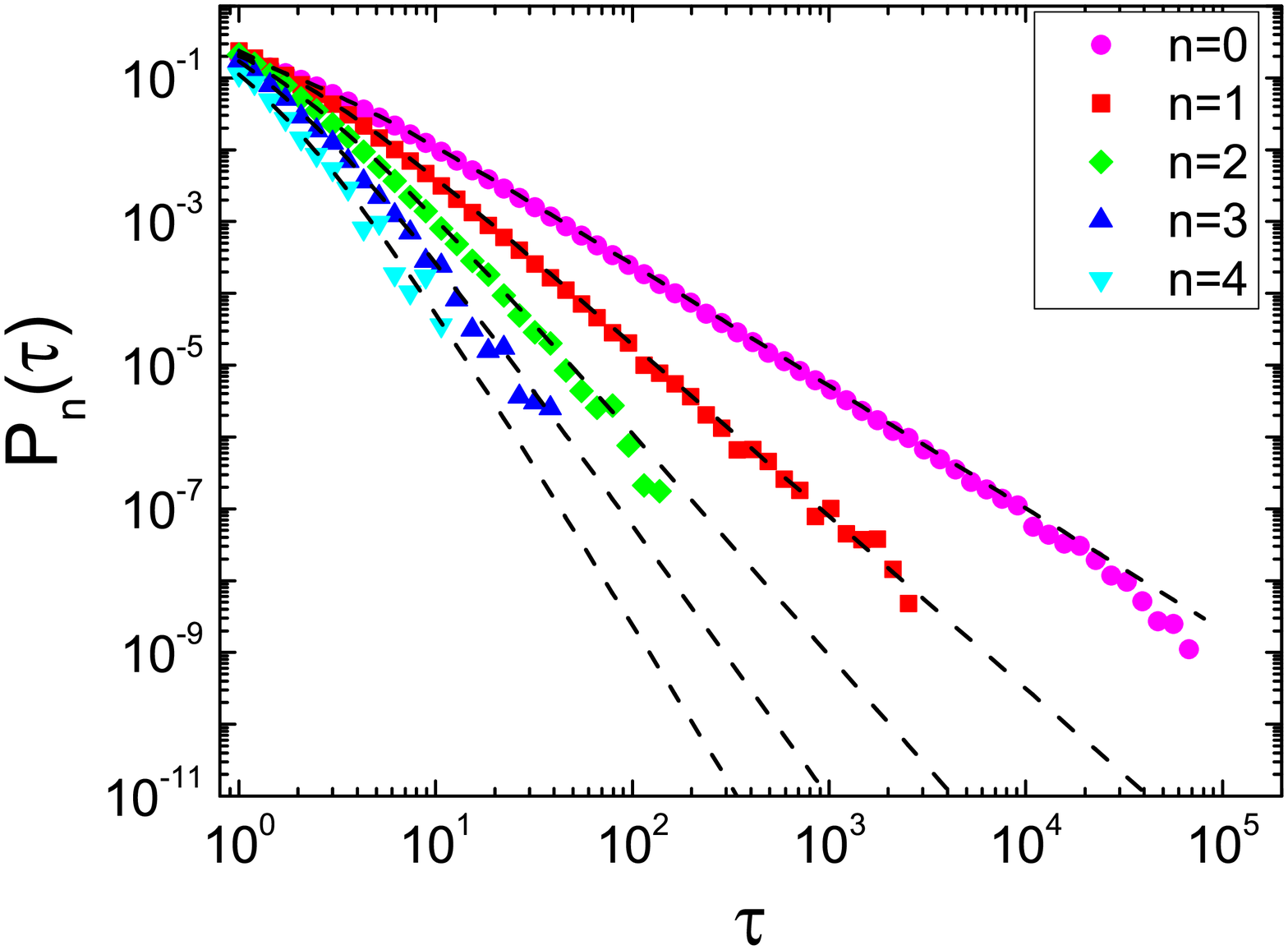}
\caption{(Color online) Distribution $P_n(\tau)$ of durations of
  groups of size $n+1$ in the non stationary region [region (II)]. The simulation is
  performed with $N=1000$ agents for a number of time steps
  $T_{max}=N\times 10^5$. The parameter used are $b_0=0.3$ and $b_1=0.7$,
  $\lambda=0.8$. The data are averaged over $10$ realizations. The
  dashed lines correspond to the analytical predictions
  Eqs. (\ref{Pn2}).}
\label{Groups_nonstationary}
\end{figure}

\begin{figure}
\includegraphics[width=0.5\textwidth]{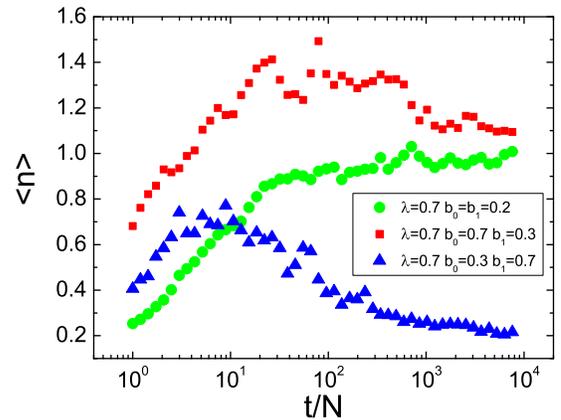}
\caption{(Color online) Average coordination number $\avg{n}$ as a function
  of time in the Region II of the phase diagram for different values
  of the parameters $\lambda$, $b_0$ and $b_1$. The data is in very good
  agreement with the theoretical expectations given by
  Eqs. $(\ref{uno})-(\ref{due})$. The simulations are performed with
  $N=1000$ agents for a number of time steps $T_{max}=N\times
  10^4$. The data are averaged over $10$ realizations.}\label{n01}
\end{figure}

The phase diagram of the model is summarized in Fig. {\ref{Fig3}}. We can distinguish between three phases.

\paragraph*{$\bullet$} {\em Region (I) -The stationary region: $b_1>0.5$,
  $b_0>(2\lambda-1)/(3\lambda-1)$ and $\lambda>0.5$-} In this region,
the self-consistent solution yields $\alpha=0$: the transition rates
$\pi_{mn}(t)$ converge rapidly to a constant value (see
Fig. \ref{p10m} for a comparison between numerics and analytics for
$\pi_{10}(t)$) and the system reaches a stationary state. In
Fig.~\ref{Groups_stationary} we compare the analytical solution given
by Eqs. (\ref{Pn2}) with the numerical simulations in the stability
region, finding perfect agreement. As predicted by Eqs. (\ref{Pn2}),
$P_n(\tau)$ decays faster as $n$ increases: larger groups are less
stable than smaller ones, as found in the empirical data sets. 
Figure \ref{Pabac} displays the distribution
$P_{AB-AC}(\tau)$ of time intervals between the start of two
consecutive contacts of a given individual, which is as well
stationary and displays a power-law behavior.

The average coordination number $\langle n \rangle$ is given by
\begin{equation}
\avg{n}=\frac{\pi_{10}}{2\lambda}\sum_{n \ge
  1}\frac{n(n+1)}{(n+1)b_1-1}\left(\frac{1-\lambda}{\lambda}\right)^{n-1} ,
\label{an}
\end{equation} 
where the detailed calculation and the value of $\pi_{10}(t)$ are
given in Appendix~\ref{A2}. This expression diverges as $\lambda\to
0.5$. In Fig. \ref{avgn} we show the perfect agreement between
the result of numerical simulations of $\avg{n}$ and the theoretical prediction.

\paragraph*{$\bullet$} {\em Region (II) -Non-stationary region: $b_1<0.5$ or
  $b_0<(2\lambda-1)/(3\lambda-1)$, and $\lambda>0.5$ -} The dynamics
in this region is non-stationary and the transition rate is decaying
with time as a power-law, as shown in Fig. \ref{p10m} where we report
$\pi_{10}(t)$ as a function of $t$. Nevertheless, the distributions of
lifetimes of groups of various sizes $P_n(\tau)$, and of inter-contact
times $P_{AB-AC}(\tau)$, remain stationary. These distributions are
shown in Figs. \ref{Groups_nonstationary} and \ref{Pabac}. 
In this region, the average
coordination number in the limit $t/N\gg 1$ remains small, even as $\lambda
\to 0.5$. In particular from the mean-field solution of the dynamics
(see appendix~\ref{A2}) the theoretical solution of the model
predicts that, for $\lambda>0.5$ and $t\rightarrow \infty$
\be
\avg{n}=1~~\mbox{for}~\alpha=1-2b_1 , 
\label{uno}
\ee
and
\be
\avg{n}=0~~\mbox{for}~\alpha=1-b_0\frac{3\lambda-1}{2\lambda-1}.
\label{due}
\ee 
Figure \ref{n01} shows the agreement of this predicted behavior
with simulation results for several values of $b_0$ and $b_1$ and
$\lambda=0.7$. In this region,  as
$\lambda\to 0.5$ with fixed $b_0$ and $b_1$, we have  $\alpha=1-2b_1$ and  $\avg{n} \to 1$.   
Therefore no diverging behavior is observed.

\begin{figure}
\includegraphics[width=0.5\textwidth]{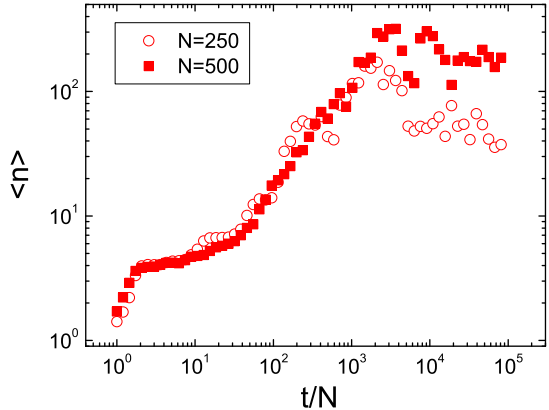}
\caption{(Color online) Average coordination number $\avg{n}$ for
  $\lambda=0.2$, $b_0=b_1=0.7$. The simulations of a single
  realization are performed with $N=250$ and $N=500$  agents, respectively, for a number of time steps $T_{max}=N
  \times 10^5$. }
\label{nvst-1}
\end{figure}

\paragraph*{$\bullet$} {\em Region (III) Strong dependence on the number of agents $N$
  and non-stationary dynamics: $\lambda<0.5$ -} In this region the
self-consistent assumption given by Eq. (\ref{pipq}) breaks down, and
we find numerically that the average coordination number $\avg{n}$
strongly depends on the number of agents $N$ and on time. In order to
give a typical example of the corresponding dynamical behavior,
Fig. \ref{nvst-1} displays $\avg{n}$ as a function of time for two
single realizations of the model corresponding to two different values
of $N$. Interestingly, the distributions of lifetimes of groups of
various sizes $P_n(\tau)$ remain stationary even in this parameter
region (not shown).

\subsection{Aggregated networks}

In the previous paragraphs, we have shown how our modeling framework
produces dynamical properties of the interactions between agents that
yield broad distributions of contact and inter-contact
times. In order to understand the structure of the resulting
interaction networks at coarser temporal resolutions, it is as well
interesting to investigate the properties of the aggregated networks,
constructed as in Sec. \ref{sec:data}. Given a starting time $t_0$ and
a temporal window $\Delta T$, the nodes of these networks are the
agents, and a link is drawn between two agents whenever they have been
in contact between $t_0$ and $t_0+\Delta T$, with a link weight given
by the total time during which they have interacted in $[t_0,t_0+\Delta T]$. As in
Sec. \ref{sec:data}, the degree $k_i$ of an agent $i$ is given by the
number of distinct agents with whom $i$ has been in contact in $[t_0,t_0+\Delta T]$, while its
strength $s_i$ is the sum of the interaction times with other agents,
and the participation ratio $Y_{2}(i)$ quantifies the heterogeneity of
the times spent by $i$ with these other agents.

\begin{figure}[tp]
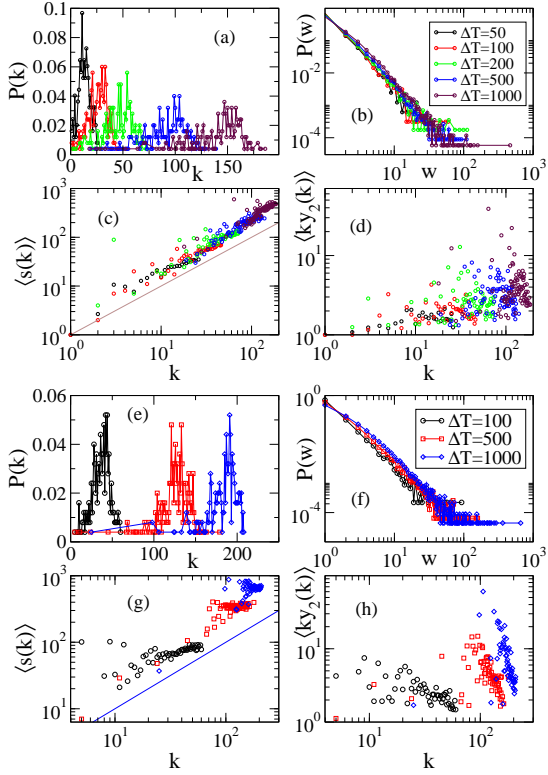

\centering
\includegraphics[width=0.4\textwidth]{fig16a}
\includegraphics[width=0.4\textwidth]{fig16b}
\caption{(Color online)
Aggregated networks' characteristics for the model with
  constant number of agents ($N=250$), for time windows $\Delta T$ of
  increasing lengths, and two sets of parameters:
  $(b_0,b_1,\lambda)=(0.55,0.8,0.9)$ (a, b, c, d) and $(0.7,0.7,0.8)$
  (e, f, g, h).}
\label{fig:aggr_nets_cstN}
\end{figure}

As an exhaustive exploration of the aggregated networks and of how
their properties depend on the model's parameter would be tedious, we
simply report in Fig.~\ref{fig:aggr_nets_cstN} the properties of
aggregated networks for increasing window lengths $\Delta T$ and for
two sets of parameters. Some properties are qualitatively similar to
the empirically observed networks. In particular, the degree
distributions are peaked around an average value that increases with
$\Delta T$. As time passes, each agent encounters more and more
distinct other agents, and the distribution $P(k)$ globally shifts
towards larger degrees. The links weights distributions are broad, and
extend to larger values as $\Delta T$ increases. Some other properties
seem to depend strongly on the model's parameters. In particular, the
average strength of nodes of degree $k$, and the average participation
ratio of nodes of degree $k$, can have shapes rather different
from the empirical ones.  Moreover, the time window lengths $\Delta T$
on which the aggregated network remains sparse are rather restricted.

\section{Extensions of the model}
\label{sec:extensions}

\subsection{Heterogeneous model}

In the previous section, we have assumed that all the agents have the
same tendency to form a group or to leave a group, that is, the
probabilities $p_n$ do not depend on the agent who performs a status
update. Real social systems display however additional complexity
since the social behavior of individuals may vary significantly across the
population. A natural extension of the model presented above consists
therefore of making the probabilities $p_n$ dependent on the agent who
is updating his/her state. To this aim, we assign to each agent $i$ a
parameter $\eta_i$ that characterizes his/her propensity to form
social interactions. In real networks this propensity will depend on
the features of the agents \cite{Bianconi:2009}.  In the model we
assume that this propensity, that we call ''sociability", is a quenched
random variable, which is assigned to each agent at the start of the
dynamical evolution and remains constant, and we assume for
simplicity that it is uniformly distributed in $[0,1]$. In this
modified model, the probability $p_n^i(t,t')$ that an agent $i$ with
coordination number $n$ since time $t'$ changes his/her coordination
number at time $t$ is given by
\begin{eqnarray}
p_0^i(t,t')&=&\frac{\eta_i}{1+(t-t')/N},\nonumber \\
p_n^i(t,t')&=&\frac{1-\eta_i}{1+(t-t')/N},\ \mbox{for}\ n\geq 1.
\label{p2}
\end{eqnarray}
In this setup, the parameters $(b_0,b_1)$, which did not depend on $i$
in Eq. (\ref{p}),
are replaced for each agent $i$ by the values $(\eta_i, 1-\eta_i)$: a large
$\eta_i$ corresponds to an agent who prefers not to be isolated.

The agents' heterogeneity adds a significant amount of complexity to
the problem, and we have reached an analytical solution of the
evolution equations only in the case of pairwise interactions
($\lambda=1$). The general case can be studied through numerical
simulations as we discuss at the end of this section.

Let us denote by $N_0(t,t',\eta)$ the number of isolated agents
with parameter $\eta_i \in [\eta,\eta+\Delta \eta]$ who have not
changed their state since time $t'$. Similarly, we indicate by
$N_1(t,t',\eta,\eta')$ the number of agents in a pair joining two
agents $i$ and $j$ with $\eta_i \in [\eta,\eta+\Delta \eta], \eta_j
\in [\eta',\eta'+\Delta \eta]$, who have been interacting since time $t'$. The
mean-field equations for the model are then given by 
\bea \frac{\partial
   N_0(t,t',\eta)}{\partial
   t}&=&-2\frac{N_0(t,t',\eta)}{N}p_0(t,t',\eta)\nonumber
 \\ &&~~~+\pi^{\eta}_{10}(t)\delta_{tt'},\nonumber \\ \frac{\partial
   N_1(t,t',\eta,\eta')}{\partial
   t}&=&-\frac{N_1(t,t',\eta,\eta')}{N}[p_1(t,t',\eta)+p_1(t,t',\eta')]\nonumber
 \\ &&+\pi^{\eta \eta'}_{01}(t)\delta_{tt'}.
\label{dynh}
\eea
With the expression for $p_n(t,t',\eta)$ given by Eqs.(\ref{p2}) we find
\bea
N_0(t,t',\eta)&=&\pi^{\eta}_{10}(t')\Big(1+\frac{t-t'}{N}\Big)^{-2\eta}, \nonumber \\
N_1(t,t',\eta,\eta')&=&\pi^{\eta\eta'}_{01}(t')\Big(1+\frac{t-t'}{N}\Big)^{-2+\eta+\eta'} .
\label{HN}
\eea 
The transition rate $\pi^\eta_{10}$ gives the rate at which agents with $\eta_i \in [\eta,\eta+\Delta \eta]$
become isolated, and $\pi^{\eta\eta'}_{01}$ is the rate at which pairs $ij$ with
$\eta_i \in [\eta,\eta+\Delta \eta], \eta_j \in [\eta',\eta'+\Delta \eta]$ are formed. These rates
can be expressed as a function 
of $N_0(t,t',\eta)$ and $N_{1}(t,t',\eta,\eta')$ according to
\bea
\pi^{\eta}_{10}(t)&=&\sum_{t',\eta'}\frac{N_1(t,t',\eta,\eta')}{N}[p_1(t,t',\eta)+p_1(t,t',\eta')],
\label{Hpi} \\
\hspace*{-5mm}
\pi^{\eta\eta'}_{01}(t)&=&2\sum_{t',t''}\frac{N_0(t,t',\eta)N_0(t,t'',\eta')}{C(t)N}p_0(t,t',\eta)p_0(t,t'',\eta'),\nonumber
\eea
where $C(t)$ is a normalization factor given by
\be
C(t)=\sum_{t^{\prime}=1}^{t}\sum_{\eta} {N_0(t,t',\eta)}p_0(t,t',\eta).
\label{Cdef}
\ee
To solve this problem with the same strategy used for the model without heterogeneity 
we make the self-consistent assumption that the transition rates are either constant 
or decaying as a power law with time:
\bea
\pi^{\eta}_{10}(t)&=&\Delta \eta\tilde{\pi}^{\eta}_{10}\Big(\frac{t}{N}\Big)^{-\alpha(\eta)}, \\
\pi^{\eta \eta'}_{01}(t)&=&\Delta\eta \Delta\eta'  \tilde{\pi}^{\eta \eta'}_{01}\Big(\frac{t}{N}\Big)^{-\alpha(\eta,\eta')}.
\eea
In appendix \ref{CC} we give the details of the self-consistent 
calculation, which leads to the analytical prediction
\bea
\alpha(\eta)=\max\left(1-2\eta, \eta-\frac{1}{2}\right), \nonumber \\
\alpha(\eta,\eta')=\alpha(\eta)+\alpha(\eta') \ ,
\eea
and the value of $\tilde{\pi}^{\eta}_{10}$ is given by 
\be
\tilde{\pi}^\eta_{10}= \left\{ \begin{array}{ll}
 \frac{\rho(\eta)}{B(1-2\eta,2\eta)} & \eta \leq \frac{1}{2} \\
\frac{\rho(\eta)}{B(\eta-\frac{1}{2},1)} & \eta \geq \frac{1}{2} 
\end{array} \right. \ .
\ee

\begin{figure}
\includegraphics[width=0.5\textwidth]{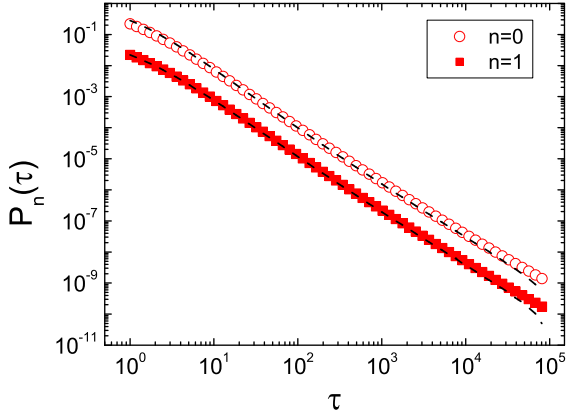}
\caption{(Color online) Distributions of times spent in state $0$
  and $1$ for the heterogeneous model. The simulation is performed with
  $N=10^4$ for a number of time steps $T_{max}=N\times 10^5$. The data
  are averaged over $10$ realizations. The symbols represent the
  simulation results (circles for $n=0$ and squares for
  $n=1$). The dashed lines represent our analytical prediction. In order to
  improve the readability of the figure we have multiplied $P_1(\tau)$
  by a factor of $10^{-1}$. }
\label{Fig4}
\end{figure}

\begin{figure}
\includegraphics[width=0.5\textwidth]{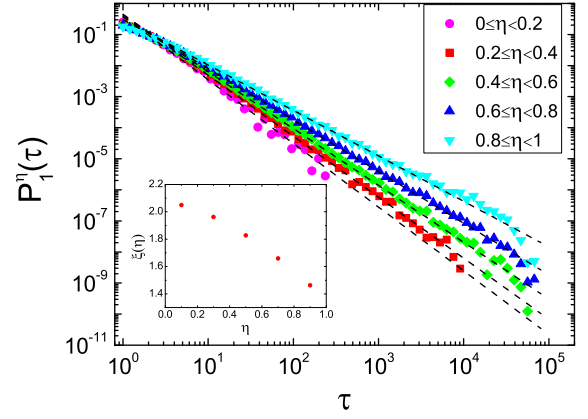}
\caption{(Color online) Distribution $P_1^{\eta}(\tau)$ of contact durations
  of individuals with sociability $\eta$ in the
  pairwise heterogeneous model. The simulations are performed with
  $N=1000$ agents and $T_{max}=N\times 10^5$ time
  steps. The data are averaged over 10 realizations. The data decays
  as a power-law $P_1^{\eta}(\tau)\propto \tau^{-\xi({\eta})}$, and 
  we report the exponents $\xi(\eta)$ as a function of $\eta$ in the inset.}
\label{Multiscaling}
\end{figure}

In order to check the validity of our mean-field calculation, we
study the probability distribution $P_0(\tau)$ of the durations of
inter-contact periods and the distribution $P_1(\tau)$ of the durations of
pairwise contacts, which are given, when averaged for a total 
simulation time $T_{max}$, by 
\bea
P_0(\tau) &\propto& \int^{T_{max}-N\tau}_{0}dt \int_0^1 d\eta
\pi^\eta_{10}(t)\eta (1+\tau)^{-2\eta-1},  \nonumber \\ 
P_1(\tau) & \propto& \int^{T_{max}-N\tau}_{0}dt \int_0^1 d\eta
\int_0^1 d\eta' \pi_{01}^{\eta\eta'} \nonumber \\
&\times&(2-\eta-\eta') (1+\tau)^{\eta+\eta'-3}, \eea
where $\rho(\eta)$ is the probability distribution of $\eta$. In
Fig. \ref{Fig4} we compare the probabilities of intercontact time
$P_0(\tau)$ and contact time $P_1(\tau)$ averaged over the full
population together with the numerical solution of the stochastic
model, showing a perfect agreement. In Fig. \ref{Multiscaling}
moreover, we show the distributions $P^{\eta}_1(\tau)$ of the contact
durations of agents with $\eta_i\in (\eta,\eta+\Delta \eta)$.
Power-law behaviors are obtained even at fixed
sociability, and the broadness of the contact duration distribution of
an agent increases with the''sociability" of the agent under
consideration.

\begin{figure}
\includegraphics[width=0.5\textwidth]{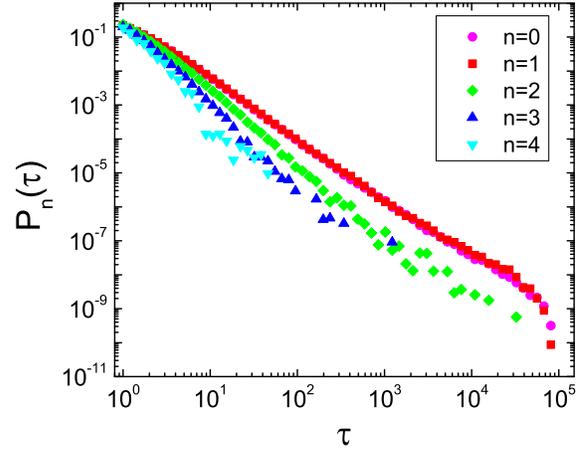}
\caption{(Color online) Distribution $P_n(\tau)$ of the durations of
  groups of size $n+1$ in the heterogeneous model with formation of
  groups of any size. The data are shown for simulations of $N=1000$
  agents performed over $T_{max}=N\times 10^5$ time steps and
  $\lambda=0.8$, averaged over $10$ realizations.}
\label{Groups2}
\end{figure}

\begin{figure}
\includegraphics[width=0.5\textwidth]{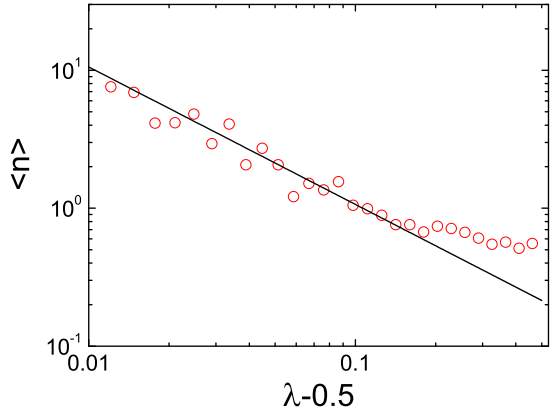}
\caption{(Color online) Average coordination number $\avg{n}$ of the
  agents in the heterogeneous case as a function of $\lambda$.  The
  solid line indicates the best fit with $\avg{n}\propto
  (\lambda-0.5)^{-\delta}$ with $\delta=0.996$ in agreement with the
  exponent $-1$ within the statistical uncertainty. The data correspond to
  simulations of $N=500$ agents performed over $T_{max}=N\times 10^3$
  time steps. The data are averaged over $10$ realizations.}
\label{n_h}
\end{figure}

As previously mentioned, the model can be extended by allowing the
formation of large groups, by setting $\lambda<1$. The results of
numerical simulations performed for a particular value of $\lambda$
are shown in Fig.\ref{Groups2}. Power law distributions of the
lifetime of groups are again found and, as in the basic model without
heterogeneity of the agents, larger groups are more unstable than
smaller groups, as $P_n(\tau)$ decays faster if the coordination
number $n$ is larger.  As the parameter $\lambda\to 0.5$ there is a
phase transition and the average coordination number diverges.  In
Fig. \ref{n_h} we show that $\avg{n}\propto (\lambda-0.5)^{-\delta}$
with $\delta=1$ within the statistical fitting error, similarly to
what happens in the homogeneous case. Overall, the main features of
the model are therefore robust with respect to the introduction of
heterogeneity in the agents' individual behavior.

\subsection{Model with variable number of agents}

When measurements on the proximity or face-to-face contacts of
individuals are performed, the number of agents present on the
premises typically fluctuates. Moreover, the activity of the agents
fluctuates as well, because for instance of day/night patterns or
periods of coffee/lunch breaks in a conference. This is in particular
the case in the data sets described in Sec. \ref{sec:data}.

In this subsection, we use these simple remarks to put forward an
extension of our model of interacting agents that mimics in a more
realistic way the data gathering process, and that can be used to
produce more realistic artificial data sets. The main point is to
consider, instead of a population with a fixed number of agents $N$,
an {\em open} system leaving the possibility for agents to leave or
enter it. To this aim, we simply introduce for each agent the
possibility to be in a state called ``absent''. In this ``absent''
state, agents may be isolated or not, but the measurement
infrastructure is not able to know it.  Statistics on the time spent
by an agent in a given state, or of the duration of contact and
inter-contact times, are thus obtained by considering only ``present''
agents. Overall, the model's dynamics is as described in
Sec. \ref{sec:model}, with parameters $b_0$, $b_1$,
$\lambda$, with the addition that at each time step, ``absent''
agents can enter the system, or agents can leave the system and become
``absent''. Various rules could be thought of, but for the sake of
simplicity we consider that an agent $i$ entering the system become isolated
($n_i \to 0$), and that agents of any state $n$ can leave the system (one
could also allow ``absent'' agents to directly enter a group, restrict
the possibility to leave to isolated agents, etc...).
In this framework, two important points have to be considered
\begin{itemize}

\item the agents who leave the system can re-enter it: the global
  number of agents (absent and present) is constant. We will see that
  a set-up in which an agent who has left the premises cannot re-enter
  the system leads to an interesting modification of the system's
  properties.

\item The decision for an agent to leave or enter the system can
  simply be given by constant probabilities. In the resulting
  dynamics, the number of agents who are present is simply fluctuating
  around a constant value that depends on these probabilities. Another
  possibility consists of {\em fixing} at each time step the number of
  present agents $N(t)$. The imposed $N(t)$ can be a given function of
  time such as for instance a periodic signal (possibly with
  superimposed stochastic noise) to mimic circadian rhythms. To mimic
  a realistic process, $N(t)$ can also be given by {\em an empirical
    time series from a real data set}, as we now consider (note that
  the total number $N$ of agents has to be at least equal to $\max_t N(t)$).
\end{itemize}

In order to impose the number of agents present in the system at each
time, we first construct the empirical time series. To this aim,
various time steps can be used. We consider the natural temporal
resolution obtained in SocioPatterns deployments, that is, $20s$, but
other resolutions could be chosen as well. Considering that agents act
independently, and make choices in a simultaneous way, we then
identify the empirical time step with one attempted status update per
present agent.  After each series of $N(t)$ attempted status updates
(according to the rules described in Sec. \ref{sec:model}), we check
if the number of present agents is smaller or larger than the desired
$N(t+1)$. If it is larger, we remove at random agents (i.e., put them
in the ``absent'' state) in order to match the desired $N(t+1)$. If it
is smaller, absent agents are introduced into the system, and put into
the isolated state. The system evolves in this way for a number of
steps equal to the number of time steps of the empirical data set, and
we monitor the time evolution of the number of contacts between
agents, and of agents in each state, as a function of time. We also
note that this model can in fact be simulated for arbitrarily long
times by simply repeating the imposed time series $N(t)$.

\begin{figure}[tp]
\centering
\includegraphics[width=0.47\textwidth]{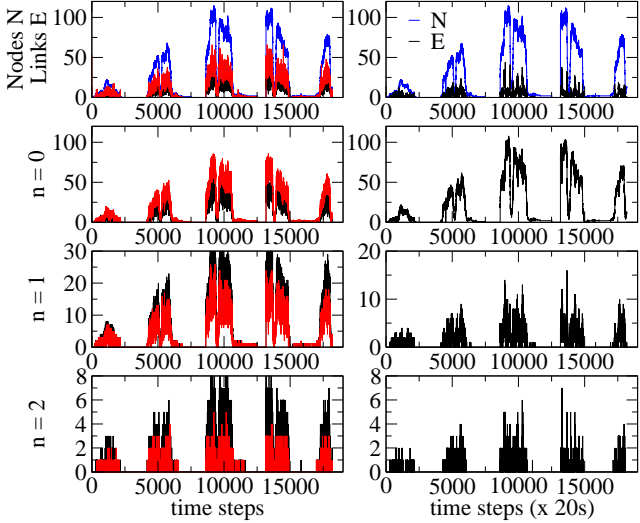}
\caption{(Color online) Timelines of the number of (from top to
  bottom): nodes and links in the instantaneous network (top),
  isolated nodes, groups of two nodes, groups of three nodes. The left
  column corresponds to the model with $N(t)$ imposed from the ESWC
  dataset and two sets of parameters, namely
  $(b_0,b_1,\lambda)=(0.55,0.8,0.9)$ (black curves) and
  $(0.7,0.7,0.8)$ (red curves), the right column to the real ESWC data
  set.}
\label{fig:timelines}
\end{figure}

\begin{figure}[tp]
\centering
\includegraphics[width=0.5\textwidth]{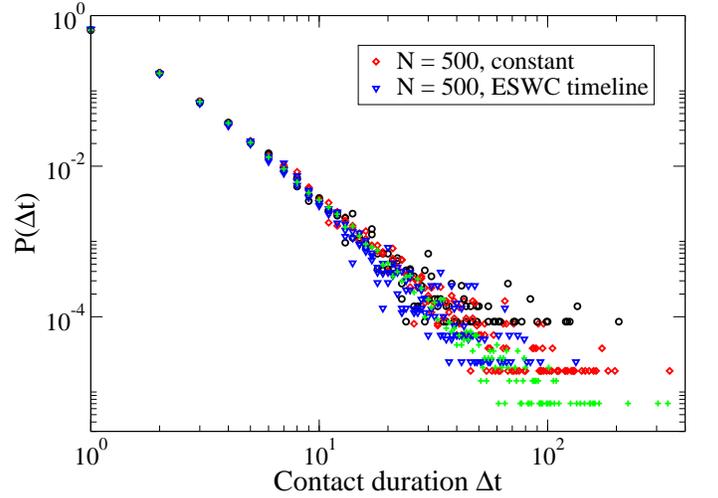}
\caption{(Color online) Distributions of contact durations, for $N(t)$
  given by the ESWC data set, and for the model with constant $N$, for
  various time windows. Parameter values are
  $(b_0,b_1,\lambda)=(0.55,0.8,0.9)$. }
\label{fig:duration1}
\end{figure}

\begin{figure}[tp]
\centering
\includegraphics[width=0.5\textwidth]{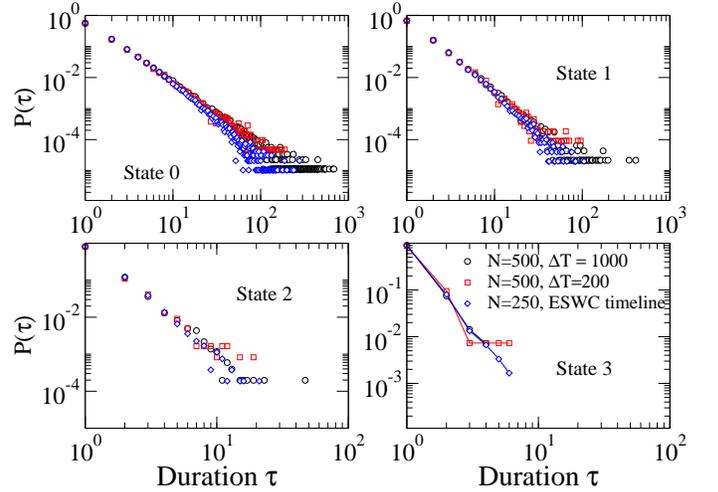}
\caption{(Color online) Distributions of the time spent in various
  states, for $N(t)$ given by the ESWC data set, and for the model
  with constant $N$.  Parameter values are
  $(b_0,b_1,\lambda)=(0.55,0.8,0.9)$. }
\label{fig:duration2}
\end{figure}

Figure \ref{fig:timelines} compares the resulting activity patterns
with the empirically monitored activity,
for two values of the parameter set $(b_0,b_1,\lambda)$, when $N(t)$
is taken from the data gathered at the scientific conference ESWC
described in Sec. \ref{sec:data}. 
We note that, although only $N(t)$ is imposed to be exactly
the same in the model and in the data, it is possible to tune the
parameters so that the other measures (number of isolated nodes, of
links, of triangles) are simultaneously similar to real
data. Figure \ref{fig:timelines} illustrates that the system's
dynamics is highly non-stationary. As shown in \cite{Cattuto:2010},
empirically observed distributions of contact durations are however
stationary. We check that this is also the case in our model by
measuring the distributions of contact durations and of time spent by
agents in each state in various time windows of different lengths,
during which $N(t)$ strongly varies. As shown in
Figs. \ref{fig:duration1} and \ref{fig:duration2}, for a given
parameter set $(b_0,b_1,\lambda)$, these distributions are broad, as
in the original model with constant number of agents, and do not depend
strongly on the imposed $N(t)$, and can be superimposed from one time
window to the next. As in the real data, the only differences come
from different cutoffs stemming from different statistics in the
different time windows.

\begin{figure}[tp]
\centering
\includegraphics[width=0.5\textwidth]{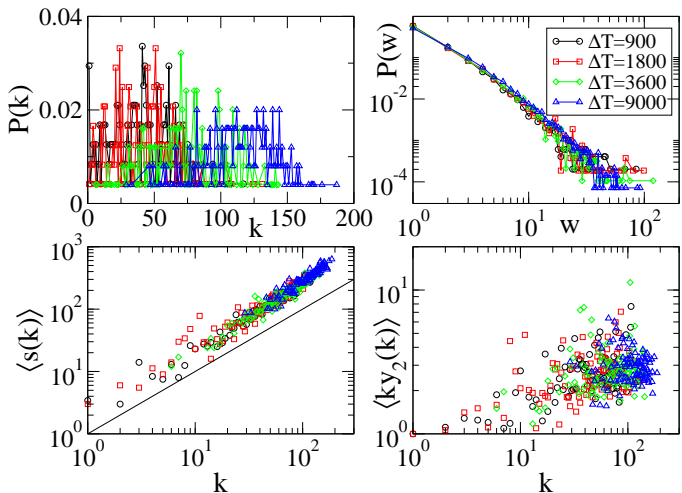}
\caption{(Color online) Aggregated networks' characteristics for
  $N(t)$ given by the ESWC time series, and $b_0=b_1=0.7$,
  $\lambda=0.8$, for various time window lengths. Top left: degree
  distributions; top right: links' weights distributions; Bottom left:
  average strength $\langle s(k) \rangle$ of nodes of degree
  $k$. Bottom right: average participation ratio $\langle k
  y_2(k)\rangle$ of nodes of degree $k$.}
\label{fig:aggr_nets_eswc_timeline}
\end{figure}

As for the study of empirical data (see Sec \ref{sec:data}),
we also construct and study the aggregated networks of contacts
between agents, on different time windows. Figure
\ref{fig:aggr_nets_eswc_timeline}, to be compared with
Fig. \ref{fig:aggr_nets_cstN} of Sec. \ref{sec:model}, illustrates the
basic properties of these networks, which are very similar to the
empirical ones shown in Fig. \ref{fig:aggr_nets_eswc} of
Sec. \ref{sec:data}. With respect to the case of constant $N$ of
Fig. \ref{fig:aggr_nets_cstN}, the degree distributions shift more
slowly toward large degrees, and remain broader, so that the average
strength of nodes of degree $k$, [$s(k)$] and the average
participation ratio of the nodes of degree $k$, [$k
y_2(k)$], keep more realistic shapes.

\begin{figure}[tp]
\centering
\includegraphics[width=0.5\textwidth]{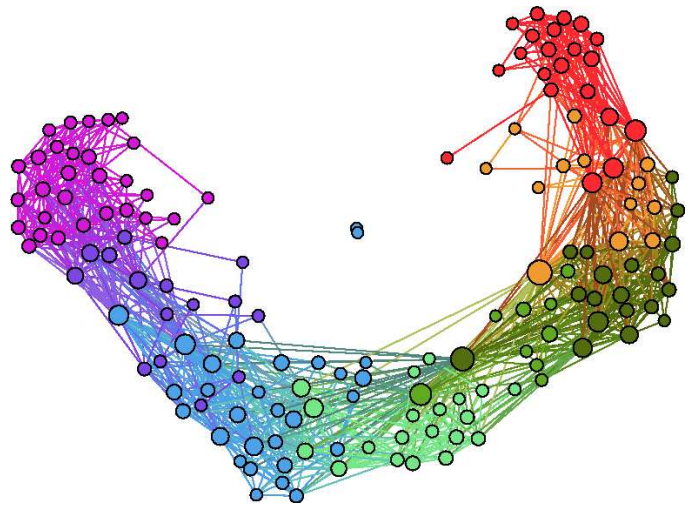}
\caption{(Color online) Example of aggregated network of $157$ nodes
  obtained by imposing the timeline of the number of agents present at
  each time during a given day of data gathered during a SocioPatterns
  deployment at the Science gallery in Dublin \cite{Isella:2011}. Here
  $(b_0,b_1,\lambda)=(0.55,0.8,0.9)$, and an agent who leaves the
  network cannot re-enter it. The nodes are colored and positioned
  according to the entry time of the corresponding visitor, as in
  \cite{Isella:2011}, from red (top right) to green (right) to blue
  (left) to violet (top left). The elongated shape of the network is
  similar to the one observed empirically in \cite{Isella:2011}.}
\label{fig:aggr_net_dublin_timeline}
\end{figure}

We finally illustrate the versatility of the modeling framework by
considering the case in which an agent who leaves the network cannot
re-enter it at future times. Such additional assumption can be
considered to model environments in which there is a flux of
individuals, such as a museum. As shown in \cite{Isella:2011}, the
daily aggregated network of the interactions between individuals is
then not a small-world, as visitors entering the museum at very
different times do not encounter each other. Figure
\ref{fig:aggr_net_dublin_timeline} exemplifies how this kind of
configurations can also be reproduced by our modeling framework, by
simply imposing the empirically measured number of visitors present on
the premises, $N(t)$, at each time step, and that an agent leaving the
system does not re-enter it. The resulting network has an elongated
shape whose topology is dictated by the timeline of the visits,
exactly as in \cite{Isella:2011}, and is strongly different from the
case in which agents can leave and re-enter the system, as in a
conference.

\section{Conclusion}
\label{sec:conclusions}

In this paper, we have studied with analytical and numerical means an
agent-based model aimed at describing the dynamics of human
interactions in social gatherings, as can be measured by recent
technological wearable sensors that measure proximity patterns of
individuals. This model is based on simple mechanisms in order to be
easily implementable and extended through the introduction of more
realistic and involved ingredients. The resulting distributions of
contact durations and of inter-contact times can be either narrowly or
broadly distributed. A detailed study of the latter case, obtained by
a mechanism reminiscent of preferential attachment, reveals moreover
interesting non-equilibrium transitions between stationary and
non-stationary phases.

We have illustrated the versatility of the model by introducing two
variations. In the first one, agents have a priori different
propensities to interact, in order to mimic the heterogeneity in the
social behaviors of individuals. We have shown that the phenomenology
of the model is then conserved, with broad distributions of contact
times, inter-contact times, and lifetimes of groups. In the second
extension of the model, the population size can fluctuate. In this
case, the population size can be taken as input of the model, as given
by an empirical time series coming from a real-world data set. We have
then shown that the model produces non-stationary dynamical networks
whose features are close to the empirically observed ones.

The present modeling framework, and in particular its extension to a
varying population, opens various perspectives. For instance, we have
considered indistinguishable agents, who enter and leave the system at
random times. It would however be possible to impose for each agent
the {\it time intervals} in which he/she is present, which could be
taken from real data. Such a procedure would produce artificial data
sets that closely mimic the empirical timelines. Although this would
happen at the cost of a rather large input of empirical information,
it would produce data sets that retain the details of the presence
properties of individual agents of the empirical data sets, but with
tunable contact and inter-contact time distributions. Another
interesting outcome of the model with varying population size is that
it makes it possible, starting from an empirical data set that is by
definition limited in time, to create a dynamical network on
arbitrarily long timescales, with the same properties as the real one,
by simply repeating the time-series $N(t)$ as many times as
required. This corresponds to an interesting way of creating a
non-stationary dynamical network, without having to repeat the {\it
  real} sequence of contacts. On each new repetition of $N(t)$, the
model will generate a new sequence of contacts.

The modeling framework we have presented, and the above-mentioned
perspectives, are particularly interesting in the perpective of the
study of dynamical processes on dynamical networks. The ability to
tune the networks' properties is indeed crucial to understanding how each
of these properties affect the dynamical processes. Generating
artificial data sets that are based on empirical ones, preserve a
certain number of their properties, modify others in a tunable way, and
can be extended to large sizes or long times, represents therefore a
very important step in such studies.

\begin{acknowledgments}
A.B. acknowledges partial support by the FET Open project DYNANETS
(number 233847) funded by the European Commission. It is a pleasure
to acknowledge the SocioPatterns project (www.sociopatterns.org), 
partially supported by the ISI Foundation, for
kindly providing the data presented in Sec. \ref{sec:data}, as well as many
interesting discussions with C. Cattuto, J.-F. Pinton and W. Van den
Broeck.
\end{acknowledgments}

\appendix
\section{Self-consistent solution of the pairwise model}
\label{AA}
In this section we give the details of the self-consistent calculation
that is able to solve for the mean-field dynamics of the pairwise
interaction model. As explained in the main text, the
rate equations Eqs. (\ref{dN01}) for this model are solved together
with the definition of the transition rates $\pi_{10}(t)$ and
$\pi_{01}(t)$ given by Eqs. (\ref{pi01}) by making the self-consistent
assumption Eq.(\ref{pi01f}).

Inserting in the definition of $\pi_{10}(t)$ and $\pi_{01}(t)$ given by Eqs. (\ref{pi01})
the structure of the solution of the mean-field dynamical Eq. (\ref{N1}) and 
the self-consistent assumption Eqs.(\ref{pi01f}), we get
\bea
\pi_{10}(t)&=&2\tilde{\pi}_{01}\frac{b_1}{N} \sum_{t'=1}^{t-1} 
\bigg(\frac{t'}{N}\bigg)^{-\alpha_1}\bigg(1+\frac{t-t'}{N}\bigg)^{-2b_1-1}.
\label{A1}
\eea
For large $N$ we can evaluate (\ref{A1}) by going to the continuous limit. Therefore in Eq. (\ref{A1}) 
we substitute the sum over time steps $t'$ with an integral over the variable $y'=t'/N$. The transition rate $\pi_{10}(y)=N\pi_{10}(t)$, that is, the average number of agents that shift from state $1\to 0$ in the unit time $y=t/N$, 
can be evaluated by the following integral:
\bea
\pi_{10}(y)&=&2N\tilde{\pi}_{01}b_1y^{-\alpha_1-2b_1}\int^{1}_{0}x^{-\alpha}(1+y^{-1}-x)^{-2b_1-1}dx \nonumber \\
&=&2N\tilde{\pi}_{01}b_1y^{-\alpha_1}f(\alpha_1,2b_1+1,y), 
\label{pi10A2}
\eea
where $f(a,b,y)$ is  given by
\be
f(a,b,y)=y^{-(b-1)}\int^{1}_{0}x^{-a}(1+y^{-1}-x)^{-b}dx.
\label{A3}
\ee
The asymptotic expansion of $f(a,b,y)$ for $y \gg 1 $ is given by
\be
f(a,b,y)=\frac{1}{b-1}+B(1-b,1-a)y^{1-b}+O\left(\frac{1}{y}+y^{-b}\right),
\label{asymf}
\ee
where $B$ is the $\beta$ function. Inserting (\ref{asymf}) into (\ref{pi10A2}) we get
\be
\pi_{10}(y)=N \tilde{\pi}_{01} y^{-\alpha_1}.
\label{A5}
\ee
This expression proves 
that the self consistent assumption given by Eq. (\ref{pi01f}) is valid. In particular since we have assumed 
\begin{equation}
\pi_{10}(y)=N\tilde{\pi}_{10} y^{-\alpha_0}\nonumber\\
\pi_{01}(y)=N\tilde{\pi}_{01} y^{-\alpha_1}
\end{equation}
these relations are consistent with the result of Eq. (\ref{A5}) obtained in the limit $N\to \infty, y\gg 1$ if
\bea
\alpha_0=\alpha_1=\alpha\nonumber \\
\tilde{\pi}_{10}=\tilde{\pi}_{01}=\tilde{\pi}.
\label{A6}
\eea
In order to find the expression for $\alpha$ and $\tilde{\pi}$ we use the conservation of 
the total number of agents. Indeed we have 
\be
\sum_{t'}\bigg[N_0(t,t')+N_1(t,t')\bigg]=N
\label{cons-N}
\ee
Using the Eqs. (\ref{N1}), (\ref{A5}) and (\ref{A6}) and substituting in Eq. (\ref{cons-N}) 
the sum over $t'$ with an integral over the variable $x=y'/y$, we get, in the limit $N\gg 1$
\bea
N\tilde{\pi} y^{-\alpha}  \bigg[ y^{-(2b_0-1)}\int^1_0 x^{-\alpha}(1+y-x)^{-2b_0}&dx& \nonumber \\
 +y^{-(2b_1-1)}\int^1_0 x^{-\alpha}(1+y-x)^{-2b_1}dx \bigg]=&N& ,
\eea
which yields
\be
\tilde{\pi} y^{-\alpha}(f(\alpha,2b_0,y)+f(\alpha,2b_1,y))=1
\ee
Finally using the asymptotic expansion  Eq. (\ref{asymf}) we get the solution given by the Eqs. 
(\ref{pitilde}) that we rewrite here for convenience
\bea
\alpha &=&\max{(0,1-2b_1,1-2b_0)}\ ,\nonumber \\
\tilde{\pi}&=&\frac{\sin{[2\pi \min{(b0,b1)}]}}{\pi}[1-\delta(\alpha,0)] \nonumber \\ 
&&+\frac{(2b_0-1)(2b_1-1)}{2(b_0+b_1-1)}\delta(\alpha,0).
\label{Apitilde}
\eea

\section{Self-consistent solution of the general model}
\label{A2}
In this appendix we solve the general model in which groups of
different size are allowed and the parameter $\lambda$ is
arbitrary. The strategy that leads to the solution of the mean-field
equation of this dynamics is essentially the same as in the
pairwise model but a new phase transition occurs when $\lambda<0.5$.
The dynamical Eqs. (\ref{dNiB}) can be solved as a function of the
variables $\pi_{mn}(t)$ by Eqs. (\ref{NiB}) and Eqs. (\ref{pig})
assuming self-consistently that that $\epsilon(t)=\hat{\epsilon}$ in
the large time limit. In order to find the analytic solution of the
mean-field dynamics it therefore important to determine the relations
between the transition rates $\pi_{mn}(t)$ and the variables
$N_n(t,t')$. These relations are given by
\bea
\pi_{1,0}(t)&=&2\lambda\sum_{t'}\frac{N_1(t,t')}{N}p_1(t,t') \nonumber \\
\pi_{n,0}(t)&=&\lambda\sum_{t'}\frac{N_n(t,t')}{N}p_1(t,t'),~n \ge 2   \nonumber \\
\pi_{n+1,n}(t)&=&(n+1)\lambda\sum_{t'}\frac{N_{n+1}(t,t')}{N}p_1(t,t'),~i \ge 1   \nonumber \\
\pi_{0,1}(t)&=&2\sum_{t'}\frac{N_0(t,t')}{N}p_0(t,t')  \nonumber \\
\pi_{0,n}(t)&=&(1-\lambda)\sum_{t'}\frac{N_{n-1}(t,t')}{N}p_1(t,t'),~n\ge 2  \nonumber \\
\pi_{n,n+1}(t)&=&(n+1)(1-\lambda)\sum_{t'}\frac{N_{n}(t,t')}{N}p_1(t,t'),~n \ge 1. \nonumber \\
\label{Bpimn}
\eea
The coupled Eqs. (\ref{NiB}), (\ref{pig}) and (\ref{Bpimn}) 
can be solved by making the additional self-consistent assumptions on the transition rates $\pi_{mn}(t)$ given by 
\bea
\pi_{mn}(y)&=&N\tilde{\pi}_{mn}y^{-\alpha_{mn}} 
\label{Ba}
\eea
where $y=t/N$ and $\pi_{mn}(y)=N\pi_{mn}(t)$.

Applying the same technique as in Appendix \ref{AA} we can prove that all 
the exponents $\alpha_{m,n}$ are equal and given by $\alpha_{m,n}=\alpha$.  
Performing straightforward calculations we get the following relations
\bea
\tilde{\pi}_{n,0}(n+1)&=&\lambda[\tilde{\pi}_{n-1,n}+\tilde{\pi}_{n+1,n}+\tilde{\pi}_{0,n}]~~\mbox{for}~n \ge 2 \nonumber \\
\tilde{\pi}_{1,0}&=&\lambda[\tilde{\pi}_{1,0}+\tilde{\pi}_{2,1}]\nonumber \\
\tilde{\pi}_{n,0}&=&(1-\lambda)\tilde{\pi}_{n-1,0}+\lambda\tilde{\pi}_{n+1,0}~~\mbox{for}~n \ge 3 \nonumber \\
\tilde{\pi}_{2,0}&=&\frac{1-\lambda}{2}\tilde{\pi}_{1,0}+\lambda\tilde{\pi}_{3,0} \nonumber\\
\tilde{\pi}_{1,0}&=&\lambda\tilde{\pi}_{1,0}+2\lambda\tilde{\pi}_{2,0}.
\label{pir}
\eea
Therefore if the self-consistent assumption is valid, the number 
of agents $N_n(t,t')$ in state $n$ since time $t'$, is given at time $t$ by
\bea
N_0(t,t')&=&\frac{\pi_{1,0}(t')}{K}\bigg(1+\frac{t-t'}{N}\bigg)^{-b_0[2+(1-\lambda)\hat{\epsilon}]} \nonumber\\
N_1(t,t')&=&\frac{\pi_{1,0}(t')}{\lambda}\bigg(1+\frac{t-t'}{N}\bigg)^{-2b_1} \nonumber\\
N_n(t,t')&=&\frac{(n+1)\pi_{n,0}(t')}{\lambda}\bigg(1+\frac{t-t_0}{N}\bigg)^{-(n+1)b_1} \,
\label{B4}
\eea
where the variable $K$ is defined by
\bea
K=\frac{\tilde{\pi}_{1,0}}{\sum_{n \ge 1}\tilde{\pi}_{n,0}}.
\label{K}
\eea
Using the relations given by Eqs. (\ref{pir}) we find 
\bea
\tilde{\pi}_{n,0}&=&\frac{1}{2}\tilde{\pi}_{1,0}\bigg(\frac{1-\lambda}{\lambda}\bigg)^{n-1}~~\mbox{for}~n \ge 2.
\label{pil}
\eea
Substituting  Eq. (\ref{pil}) in the definition of $K$, Eq. (\ref{K}), we find that $K$ is only defined for $\lambda>0.5$.
For $\lambda<0.5$ the summation in Eq. (\ref{K}) is in fact divergent and there is a breakdown 
of the self-consistent assumption Eq. (\ref{Ba}).
For $\lambda>0.5$ we can perform the summation and we get 
\bea
K&=&\frac{2(2\lambda-1)}{3\lambda-1} \ ,\nonumber\\
\hat{\epsilon}&=&\frac{1}{2\lambda-1} \ .
\eea
Finally the value of $\alpha$ and $\tilde{\pi}_{1,0}$ are found by enforcing the conservation law of the number of agent $N$
\be
\sum_{t'=1}^t\sum_{n}N_n(t,t')=N .
\ee
Therefore, in the large   $y$ limit $y \gg1$ we get the solution
\be 
\alpha=\max\left(0, 1-b_0\frac{3\lambda-1}{2\lambda-1}, 1-2b_1\right) .
\label{ag2}
\ee
The  value of  $\tilde{\pi}_{1,0}$ depends on the value assumed by  $\alpha$.
\begin{itemize}
\item[(1)]For $\alpha=0$, the value of $\tilde{\pi}_{1,0}$ is given by 
\bea
\tilde{\pi}_{1,0}&=&\bigg[\frac{1}{2(b_0-\frac{2\lambda-1}{3\lambda-1})} \nonumber \\
&&+\frac{1}{2\lambda}\sum_{n \ge 1}\frac{n+1}{(n+1)b_1-1}\left(\frac{1-\lambda}{\lambda}\right)^{n-1}\bigg]^{-1} .
\label{pig1}
\eea
\item[(2)]For $\alpha=1-b_0\frac{3\lambda-1}{2\lambda-1}$, the value of $\tilde{\pi}_{1,0}$  is given by
\be
\tilde{\pi}_{1,0}=\frac{2(2\lambda-1)}{3\lambda-1}\frac{1}{B(1-b_0\frac{3\lambda-1}{2\lambda-1},b_0\frac{3\lambda-1}{2\lambda-1})},
\label{pig2}
\ee
where $B(a,b)$ indicates  the Beta function. 
\item[(3)]
For $\alpha=1-2b_1$, the value of $\tilde{\pi}_{1,0}$ is given by 
\be
\tilde{\pi}_{1,0}=\frac{\lambda}{B(1-2b_1,2b_1)}
\label{pig3}
\ee
where $B(a,b)$  indicates  the Beta function. 
\end{itemize}
The average coordination number is defined by
\be
\avg{n}=\sum_{t'}^{t}\sum_{n=0}^N nN_n(t,t').
\label{ng1}
\ee
Substituting Eqs. (\ref{B4}) to the definition of $\avg{n}$, Eq. (\ref{ng1}) and applying the same transformation in Eq. (\ref{pi10A2}) to evaluate the integral over $t$, we get
\be
\avg{n}=\sum_{n=1}^N \frac{\tilde{\pi}_{n0}}{\lambda}(n+1-\delta_{n,1})y^{-\alpha} f(\alpha,(n+1)b_1,y)
\label{ng2}
\ee
where $y=t/N$ and $f(a,b,y)$ is defined in Eq. (\ref{A3}). Substituting the asymptotic expansion Eq. (\ref{asymf}) into (\ref{ng2}), we get
\bea
\avg{n}&=&\sum_{n=1}^N \frac{\tilde{\pi}_{n0}}{\lambda}(n+1-\delta_{n,1})y^{-\alpha} \bigg[\frac{1}{(n+1)b_1-1} \nonumber \\
&+&B(1-(n+1)b_1,1-\alpha)y^{1-(n+1)b_1}\bigg].
\label{ng3}
\eea
where $B(a,b)$ indicates the Beta function. In the asymtotic limit $y \rightarrow \infty$, using Eqs. (\ref{pil}), (\ref{pig1})-(\ref{pig3}) and counting only the leading terms in Eq. (\ref{ng3}) to compute $\avg{n}$ for different value of $\alpha$,  we can recover Eqs. (\ref{an})-(\ref{due}).

\section{Self-consistent solution of the heterogeneous model for $\lambda=1$}
\label{CC}
In this appendix we show the self-consistent  calculations that solve analytically the heterogeneous model with 
pairwise interactions.
 
We assume self-consistently that the transition rate $\pi_{10}^{\eta}(t)$ and $\pi_{01}^{\eta,\eta'}$ 
decay in time as a power-law, i.e. we assume
\bea
\pi^{\eta}_{10}(t)&=&\Delta \eta\tilde{\pi}^{\eta}_{10}\Big(\frac{t}{N}\Big)^{-\alpha(\eta)} \nonumber\\
\pi^{\eta \eta'}_{01}(t)&=&\Delta \eta \Delta \eta' \tilde{\pi}^{\eta \eta'}_{01}\Big(\frac{t}{N}\Big)^{-\alpha(\eta,\eta')}
\label{Assum}
\eea
Inserting this self-consistent assumption and the structure of the solution given by 
Eqs. (\ref{HN}) in   Eqs. (\ref{Hpi}) we can evaluate $\tilde{\pi}^{\eta,\eta'}$ in the limit $N\to \infty$. 
Therefore we get,
\bea
\tilde{\pi}^{\eta\eta'}_{01}&y&^{-\alpha(\eta,\eta')}=\frac{2N}{C(y)}\eta\tilde{\pi}^{\eta}_{10}y^{-\alpha(\eta)}f(\alpha(\eta),2\eta+1,y) \nonumber \\
&&\eta'\tilde{\pi}^{\eta'}_{10}y^{-\alpha(\eta')}f(\alpha(\eta'),2\eta'+1,y) 
\label{Cpie}
\eea
where $f(a,b, y)$ is given by 
\be 
f(a,b,y)=y^{-(b-1)}\int^{1}_{0}x^{-a}(1+y^{-1}-x)^{-b}dx.
\ee
The asymptotic expansion to $f(a,b,y)$ for $y \gg1 $ is given by 
\be
f(a,b,y)=\frac{1}{b-1}+B(1-b,1-a)y^{1-b}+O\left(\frac{1}{y}+y^{-b}\right)
\label{Casym}
\ee
where $B$ is the Beta function. Inserting (\ref{Casym}) into (\ref{Cpie}), we get in the limit $y\gg1$ 
\be
\tilde{\pi}^{\eta \eta'}_{01}y^{-\alpha(\eta,\eta')}=\frac{N}{2C(y)}\tilde{\pi}^{\eta}_{10}y^{-\alpha_0(\eta)}\tilde{\pi}^{\eta'}_{10}y^{-\alpha(\eta')}
\label{Cpieep}
\ee
Similarly, inserting (\ref{Assum}) into the definition of $C(y)$ given by Eq. (\ref{Cdef}) we get, in the limit $y\gg1$
\be
C(y)=\frac{N}{2}\int^1_0 y^{-\alpha(\eta)}\tilde{\pi}^{\eta}_{10}d\eta
\ee
where we make use of the asymptotic expansion (\ref{Casym}).
 In the  limit $y \gg1  $, the integral above can be calculated approximately by the saddle point method if $\tilde{\pi}^{\eta}_{10}$ changes with $\eta $ much slower than $y^{-\alpha(\eta)}$. Therefore  we have 
 \be
\frac{2C(y)}{N}=\tilde{\pi}^{\eta^{\star}}_{10}y^{-\gamma}
\label{CCt}
\ee
where $\gamma $ and $\eta^{\star}$ are  given by 
\bea
\gamma &=&\min_\eta\alpha(\eta) \nonumber \\
\eta^{\star} &=& \mbox{argmin}_{\eta} \alpha(\eta).
\eea
By comparing both sides of Eq. (\ref{Cpieep})  and using Eq. (\ref{CCt}) we get
\bea
\tilde{\pi}^{\eta \eta'}_{01}&=&\frac{1}{\tilde{\pi}^{\eta^{\star}}_{10}}\tilde{\pi}^{\eta}_{10}\tilde{\pi}^{\eta'}_{10}\nonumber \\
\alpha(\eta,\eta')&=&\alpha(\eta)+\alpha(\eta')+\gamma.
\eea
Finally, in order to fully solve the problem we impose the conservation laws of this heterogeneous model.
In particular the total number of agent with value $\eta_i\in(\eta,\eta+\Delta \eta)$ is given  by the following relation, 
\be
\sum_{t'}[N_0(t,t',\eta)+\sum_{\eta'}N_1(t,t',\eta,\eta')]=N\Delta(\eta).
\label{Cc}
\ee

Inserting the self-consistent anzatz Eq. (\ref{Assum}) for $\pi_{01}(t)$ and Eq. (\ref{Cpieep}) into Eq. (\ref{Cc})  we get, in the continuous limit approximation valid for $N\gg1$,
\bea
\tilde{\pi}^\eta_{10}y^{-\alpha(\eta)}&=&\bigg[ \frac{\theta(2\eta-1)}{2\eta-1}+\theta(1-2\eta)  \nonumber \\  
&& \hspace{-25mm}\times B(1-2\eta, 1-\alpha(\eta))y^{1-2\eta}+I(\eta) \bigg]^{-1}
\eea
where
\bea
I(\eta)&=&\frac{N}{2C(y)} \int^1_0 \bigg[ \frac{\theta(1-\eta-\eta')}{1-\eta-\eta'}+\theta(\eta+\eta'-1) \nonumber \\
&\times&B(\eta+\eta'-1,1-\alpha(\eta'))y^{\eta+\eta'-1} \bigg]  \nonumber \\ 
&\times& \pi^{\eta'}_{10}y^{-\alpha(\eta')}d\eta'.
\label{CI}
\eea
We compute $I(\eta)$ defined in Eq. (\ref{CI}) by counting the leading term only.
Therefore we find 
\be
\alpha(\eta)=\max(0, 1-2\eta, \eta-1+\gamma+D) 
\label{Ca}
\ee
with $D$ given by 
\be
D=\max_\eta[\eta-\alpha(\eta)]. 
\label{CD}
\ee
Solving the  Eqs. (\ref{Ca})  and (\ref{CD}) we get $\gamma=0$ and $D=\frac{1}{2}$ and $\eta^{\star}=1/2$.
Therefore we can determine the exponent $\alpha(\eta)$ and $\alpha(\eta,\eta')$ that are given by 
\bea
\alpha(\eta)&=&\max\left(1-2\eta, \eta-\frac{1}{2}\right) \nonumber \\
\alpha(\eta,\eta')&=&\alpha(\eta)+\alpha(\eta').
\eea
Moreover the constants $\tilde{\pi}_{10}^{\eta}$ are given, in the limit $N\gg1$ and $y\gg1$, by
\be
\tilde{\pi}^{\eta}_{10}= \left\{ \begin{array}{ll}
\frac{\rho(\eta)}{B(1-2\eta,2\eta)} & \eta \leq \frac{1}{2} \\
\frac{\rho(\eta)}{B(\eta-\frac{1}{2},1)} & \eta \geq \frac{1}{2} .
\end{array} \right.
\ee

Solving equation (\ref{C18}), let $\gamma+D \leq \frac{1}{2} $, then
\be
\alpha(\eta)= \left\{ \begin{array}{ll}
 1-2\eta & \eta \leq \frac{1}{2} \\
0 & \frac{1}{2} \leq \eta \leq 1-\gamma-D \\
\eta-1+\gamma+D & \eta \geq 1-\gamma-D
\end{array} \right. 
\ee
and
\be
 \eta-\alpha(\eta)= \left\{ \begin{array}{ll}
 3\eta-1 & \eta \leq \frac{1}{2} \\
\eta & \frac{1}{2} \leq \eta \leq 1-\gamma-D \\
1-\gamma-D & \eta \geq 1-\gamma-D
\end{array} \right.
\label{C18} 
\ee
obviously, $\gamma=0$
and D is reached either at $\eta=\frac{1}{2}$ or $\eta=1-\gamma-D$, so
\be
D=\max(\frac{1}{2},1-D)
\ee
The only solution to the above expression is
$D=\frac{1}{2}$.
Similarly, for $\gamma+D \geq \frac{1}{2}$,
\be
 \alpha(\eta)= \left\{ \begin{array}{ll}
 1-2\eta & \eta \leq \frac{2-\gamma-D}{3} \\
\eta-1+\gamma+D & \eta \geq \frac{2-\gamma-D}{3}
\end{array} \right. 
\ee
\be
 \eta-\alpha(\eta)= \left\{ \begin{array}{ll}
 3\eta-1 & \eta \leq \frac{2-\gamma-D}{3} \\
1-\gamma-D & \eta \geq \frac{2-\gamma-D}{3}
\end{array} \right. 
\ee
Both $\gamma$ and D are reached at $\eta=\frac{2-\gamma-D}{3}$, so
\be
 \left\{ \begin{array}{l}
 \gamma=1-\frac{2(2-\gamma-D)}{3}  \\
D=(2-\gamma-D)-1 
\end{array} \right. 
\ee

\end{document}